\documentclass[prd,twocolumn,showpacs,superscriptaddress]{revtex4}
\usepackage{amsfonts}
\usepackage{amsmath}
\usepackage{amssymb}
\usepackage{bm}
\usepackage{dcolumn}
\usepackage{epsfig}
\usepackage{graphicx}
\usepackage{graphics}
\usepackage[latin1]{inputenc}
\usepackage{latexsym}
\usepackage{rotating}
\usepackage{hyperref}

\newcommand\be{\begin{equation}}
\newcommand\ba{\begin{eqnarray}}
\newcommand\ee{\end{equation}}
\newcommand\ea{\end{eqnarray}}

\newcommand{\mb}[1]{\mbox{\boldmath $#1$}}

\def \nn   {\nonumber}
\def \met  {\mbox{g}}

\newcommand{\GW}{{\mbox{\tiny GW}}}
\newcommand{\ZM}{{\mbox{\tiny ZM}}}

\newcommand{\CPM}{{\mbox{\tiny CPM}}}

\newcommand{\AXIAL}{{\mbox{\tiny Axial}}}
\newcommand{\POLAR}{{\mbox{\tiny Polar}}}
\newcommand{\pont}{{\,^\ast\!}R\,R}

\begin{document}
\title{Perturbations of Schwarzschild Black Holes in Chern-Simons
  Modified Gravity}

\author{Nicol\'as Yunes}
\affiliation{Center for Gravitational Wave Physics and Institute for
Gravitational Physics and Geometry, \\ Department of Physics, 
The Pennsylvania State University, University Park, PA 16802, USA}

\author{Carlos F.~Sopuerta}
\affiliation{Department of Physics, University of Guelph, Ontario N1G 2W1, Canada}

\affiliation{Institut de Ci\`encies de l'Espai (CSIC), 
Facultat de Ci\`encies, Campus UAB, Torre C5 parells, 
E-08193 Bellaterra, Spain}

\affiliation{Institut d'Estudis Espacials de Catalunya (IEEC), Ed.~Nexus-201,
Gran Capit\`a 2-4, E-08034 Barcelona, Spain}

\date{\today}

\preprint{IGC-07/12-3}

%%%%%%%%%%%%%%%%%%%%%%%%%%%%%%%%%%%%%%%%%%%%%%%%%%%%%%%%%%%%%%%%%%%%%%%%%%%%%%
\begin{abstract}
  
  We study perturbations of a Schwarzschild black hole in Chern-Simons
  modified gravity.  We begin by showing that Birkhoff's theorem holds
  for a wide family of Chern-Simons {\em coupling} functions, a scalar
  field present in the theory that controls the strength of the
  Chern-Simons correction to the Einstein-Hilbert action. After
  decomposing the perturbations in spherical harmonics, we study the
  linearized modified field equations and find that axial and polar
  modes are coupled, in contrast to general relativity. The divergence
  of the modified equations leads to the {\em Pontryagin constraint},
  which forces the vanishing of the Cunningham-Price-Moncrief master
  function associated with axial modes. We analyze the structure of
  these equations and find that the appearance of the Pontryagin
  constraint yields an overconstrained system that does not allow for
  generic black hole oscillations. We illustrate this situation by
  studying the case characterized by a canonical choice of the
  coupling function and pure-parity perturbative modes.  We end with a
  discussion of how to extend Chern-Simons modified gravity to bypass
  the Pontryagin constraint and the suppression of perturbations.

\end{abstract}


\maketitle

%%%%%%%%%%%%%%%%%%%%%%%%%%%%%%%%%%%%%%%%%%%%%%%%%%%%%%%%%%%%%%%%%%%%%%%%%%%%%%
\section{Introduction}
\label{intro}

Extensions of general relativity (GR) are inherently interesting
because they hold the promise to address unresolved problems in
cosmology and astrophysics.  One such extension is Chern-Simons (CS)
modified gravity~\cite{jackiw:2003:cmo}, which has recently been used
to propose an explanation to the cosmic baryon
asymmetry~\cite{alexander:2004:lfg} and polarization in the cosmic
microwave background (CMB)~\cite{Lue:1998mq}. This extension has also
attracted some recent interest because it might allow for a test of a
wide class of string theories~\cite{Alexander:2007:bgw}.

CS modified gravity introduces a well-motivated correction to the
Einstein-Hilbert action: the product of a parity-violating,
{\emph{Pontryagin term}}~\footnote{See~\cite{Grumiller:2007rv} for a
  discussion of this term and some physical interpretations} and a
scalar field or {\emph{CS coupling function}} that controls the
strength of the correction. Such a modification has deep roots in the
standard model, since chiral, gauge and gravitational anomalies
possess Pontryagin-like structures. CS modified gravity is also
motivated by string theory and loop quantum gravity~\cite{stephon}. In
the former, it arises through the Green-Schwarz
mechanism~\cite{Green:1987mn}, as an anomaly-canceling term. In fact,
the CS term is a requirement of all model-independent extension of
$4$-dimensional compactifications of string
theory~\cite{Polchinski:1998rr}.

Some theoretical aspects of CS modified gravity have recently been
investigated. Cosmological studies have been carried out
in~\cite{Lue:1998mq,Li:2006ss,Alexander:2006mt} in connection with the
Cosmic Microwave Background radiation and
in~\cite{alexander:2004:lfg,Alexander:2004xd} in the context of
leptogenesis. Gravitational wave solutions have also been studied
in~\cite{jackiw:2003:cmo,Alexander:2004wk,Alexander:2007zg,Alexander:2007vt},
where they were found to possess amplitude birefringence, possibly
leading to a test of the theory~\cite{Alexander:2007:bgw}. Weak-field
solutions for spinning objects in CS modified gravity have been
studied in~\cite{Alexander:2007zg,Alexander:2007vt}, leading to a
prediction of gyromagnetic precession that differs from GR, and later
to a constraint on the magnitude of the CS
correction~\cite{Smith:2007jm}. Further discussion on these issues and
others related can be found
in~\cite{Kostelecky:2003fs,Mariz:2004cv,Alexander:2004wk,Bluhm:2004ep,Eling:2004dk,Lyth:2005jf,Mattingly:2005re,Lehnert:2006rp,Alexander:2006mt,Hariton:2006zj,Alexander:2007qe,Guarrera:2007tu,Alexander:2007vt,Konno:2007ze,Smith:2007jm,Fischler:2007tj,Tekin:2007rn,Grumiller:2007gm}
and references therein.

Solutions to the CS modified field equations that represent
interesting physical configurations have also attracted attention.
Apart from first formalizing the theory, Jackiw and
Pi~\cite{jackiw:2003:cmo} also showed that the Schwarzschild and
gravitational plane-wave spacetimes remain solutions in CS modified
gravity.  The Reissner-Nordstrom and Friedman-Robertson-Walker metrics
were also found to persist in the theory~\cite{Guarrera:2007tu}. On
the perturbative front,~\cite{Alexander:2007zg,Alexander:2007vt} found
a weak-field solution for orbiting, spinning objects,
while~\cite{Konno:2007ze} showed that the weak-field limit of the Kerr
metric remained a CS solution to first order in the metric
perturbation for a specific choice non-canonical of the CS coupling.
Exact solutions that represent spinning objects in CS modified gravity
have so far only been studied in~\cite{Grumiller:2007rv}, where it was
suggested that these exist provided the gravitomagnetic sector of the
metric is non-vanishing or that stationarity is broken.

In this paper, we study perturbations of a Schwarzschild black hole
(BH) in CS modified gravity~\footnote{The term``CS modified gravity''
  stands for the formulation introduced in~\cite{jackiw:2003:cmo}. Extensions of
  this formulation have been considered
  in~\cite{Smith:2007jm,Grumiller:2007rv} and we shall discuss how
  these affect the analysis of this paper in the last section.}. We
begin by showing that Birkhoff's theorem --the statement that the
Schwarzschild solution is the only vacuum, spherically-symmetric
solution of the theory-- persists in the modified theory for a wide
class of coupling functions.  After decomposing the metric
perturbations in spherical harmonics, we find equations that govern the
behaviour of each harmonic. In GR, the divergence of the field
equations (both at the perturbative and non-perturbative levels)
implies energy-momentum conservation and, in the absence of matter, it
becomes an identity that ensures diffeomorphism invariance. In CS
gravity, however, the divergence of the field equations (again at the
perturbative and non-perturbative levels) leads to a constraint, the
so-called Pontryagin constraint, which ensures that the
theory remains diffeomorphic invariant and that the Strong Equivalence
Principle holds. This constraint imposes a restriction on the class of
metrics that can be solutions of the theory, specially in
vacuum~\cite{Grumiller:2007rv}.  At the perturbative level, the
Pontryagin constraint is automatically satisfied for a flat
background, but we show that for a BH background it forces the
Cunningham-Price-Moncrief (CPM) master function, associated with axial
perturbations (perturbations of odd parity), to vanish.  Such a strong
restriction on the possible perturbative modes a BH is allowed to
possess is a distinctive feature of CS gravity.

After investigating these preliminary issues, we concentrate on the
study of general perturbations of a Schwarzschild
spacetime~\footnote{BH perturbations were previously studied
  in~\cite{Konno:2007ze}, but only perturbative modes that resemble
  the weak-field limit of the Kerr metric were considered. In this
  paper, we consider the most general perturbations in a Schwarzschild
  background.}.  We find that the CS modification introduces new terms
into the field equations that results in a mixing of polar and axial
parity modes.  Therefore, in CS modified gravity modes with different
parity do not decouple, as is the case in GR.  Moreover, the new terms
contain third-order derivatives that change the basic structure of the
partial differential equations that describe the behavior of the
metric perturbations. In particular, the Pontryagin constraint
constitutes a new equation for the metric perturbations, which in
general leads to an overdetermined system.  A priori, it remains
unclear whether this system has non-trivial solutions,
{\emph{i.e.}}~whether the entire set of field equations is compatible.
This paper shows that at least a wide class of BH oscillation modes
are forbidden in the modified theory for a large family of CS coupling
functions.  More specifically, we show that it is not possible to have
either pure axial or pure polar oscillations of a BH in CS gravity for
a wide class of coupling functions.

We end with a discussion of possible routes to extend CS modified
gravity such that generic BH oscillations are allowed.  A promising
route is to allow the CS coupling function to be a dynamical quantity.
In this case, the Pontryagin constraint is balanced by the equation of
motion of the scalar field, thus preventing the CPM function to
identically vanish. Moreover, the field equations are modified by the
introduction of a stress-energy tensor for the CS scalar field and by
terms describing perturbations of the scalar field.  The extended set
of field equations, which has a certain ambiguity encoded in the
potential of the CS scalar field, is no longer overdetermined, and
hence, it provides a suitable framework to study the modified dynamics
of BH oscillations.

This paper is divided as follows: Sec.~\ref{CS-basics} describes the
basics of CS modified gravity; Sec.~\ref{birk} establishes Birkhoff's
theorem for a family of CS coupling functions; Sec.~\ref{BHPT} begins
by describing the basics of Schwarzschild BH perturbation theory, and
then discusses the consequences of the linearized Pontryagin
constraint and the structure of the modified field equations;
Sec.~\ref{BHPT-canon} investigates perturbations of canonical CS
gravity, including the impossibility of having either purely polar or
purely axial perturbative modes; Sec.~\ref{beyond} points to
directions in which the theory could be extended to bypass the
Pontryagin constraint and allow for generic oscillations of a
Schwarzschild BH; Sec.~\ref{conclusions} concludes and points to
future work.

We use the following conventions throughout this work. Greek letters
and a semicolon are used to denote indices and covariant
differentiation respectively on the $4$-dimensional spacetime.
In some cases, we denote covariant differentiation with respect to a 
Schwarzschild background metric by $\bar\nabla^{}_\mu$.
Partial differentiation of a quantity $Q$ with respect to the
coordinate $x^{\mu}$ is denoted as $\partial^{}_{x^{\mu}}Q$ or
$Q^{}_{,\mu}$.
Symmetrization and antisymmetrization is denoted with parenthesis and
square brackets around the indices respectively, such as $A^{}_{(ab)}
:= [A^{}_{ab} + A^{}_{ba}]/2$ and $A^{}_{[ab]} := [A^{}_{ab} -
A^{}_{ba}]/2$. We use the metric signature $(-,+,+,+)$ and geometric
units in which $G = c = 1$.

%%%%%%%%%%%%%%%%%%%%%%%%%%%%%%%%%%%%%%%%%%%%%%%%%%%%%%%%%%%%%%%%%%%%%%%%%%%%%%
\section{Introduction to Chern-Simons Modified Gravity}
\label{CS-basics}

In this section we describe the basics of CS modified gravity as it
was introduced by Jackiw and Pi~\cite{jackiw:2003:cmo}.  (more details
can be found in~\cite{jackiw:2003:cmo,alexander:2004:lfg,Grumiller:2007rv} and
references therein). In this paper, we shall be mainly concerned with
this formulation~\cite{jackiw:2003:cmo}, but we will also discuss
possible extensions later on.

The CS extension of the GR action is given by~\cite{Smith:2007jm}
\ba
\label{full-action}
S^{}_{\rm{CS}} &:=& \kappa \int d^4x  \sqrt{-\met} \left[ R +
{\cal{L}}_{\rm{mat}} - \frac{1}{4} \theta \pont \right], 
\ea
where $\kappa = 1/(16 \pi)$, $\met$ is the determinant of the
spacetime metric $\met^{}_{\mu\nu}$, $R$ stands for the Ricci scalar and
${\cal{L}}^{}_{\rm{mat}}$ is the Lagrangian density of the matter
fields.  We recognize the first two terms as the Einstein-Hilbert
action in the presence of matter, while the last term is the CS
modification, defined via
\be
\pont  := R^{}_{\alpha \beta \gamma \delta}
{\,^\ast\!}R^{\alpha \beta \gamma \delta} =
\frac{1}{2} R^{}_{\alpha \beta \gamma \delta}
\eta^{\alpha \beta \mu \nu} R^{\gamma \delta}{}^{}_{\mu \nu}\,,
\label{CSmodification}
\ee
where the asterisk denotes the dual tensor, constructed using the
$4$-dimensional Levi-Civita tensor $\eta^{\alpha \beta \mu
  \nu}$~\footnote{In non-vacuum spacetimes, the dual of the Riemann
  tensor with respect to the first pair of indices does not coincide
  with the dual with respect to the second pair of indices. In this
  paper we use the second type of dual, but the results of this paper
  are independent of this choice. This can also be seen from the
  condition $\pont = {\,^\ast\!}C C$~\cite{danielpriv}}.  The strength
of the CS correction is controlled by the scalar field $\theta$, which
we shall refer to as the {\emph{CS coupling function}} or {\emph{CS
    scalar field}}.  The so-called {\em canonical} choice of $\theta$
corresponds to~\cite{jackiw:2003:cmo}
\be
\label{can}
\theta_{\rm{can}} := [t/\mu,0,0,0]\,,
\ee
where $\mu$ is a dimensional parameter.

As usual, we obtain the (modified) field equations by varying the action
with respect to the metric, yielding 
\ba
\label{EEs}
G^{}_{\mu \nu} + C^{}_{\mu \nu}  &=& 8 \pi T_{\mu \nu}^{\rm{mat}},
\ea
where $T_{\mu \nu}^{\rm{mat}}$ is the matter stress-energy tensor.
We shall refer to the quantity $C^{}_{\mu \nu}$ as the C-tensor, defined
as 
\be
\label{C-def}
C_{\mu \nu} := C_{\mu \nu}^{(1)} + C_{\mu \nu}^{(2)}, 
\ee
where 
\ba
\label{cspart1}
C_{\mu \nu}^{(1)} &:=& \theta^{}_{;\sigma}
  \eta^{\sigma}{}_{(\mu}{}^{\alpha \beta} R^{}_{\nu) \beta;\alpha} \,,
\\
\label{cspart2}
C_{\mu \nu}^{(2)} &=& \theta^{}_{;\sigma \tau}
{\,^\ast\!}R^{\sigma}{}_{(\mu}{}^{\tau}{}_{\nu)} \,. 
\ea
The spatial sector of the C-tensor reduces to the $3$-dimensional
Cotton tensor in some symmetric
cases~\footnote{In~\cite{jackiw:2003:cmo}, $C^{}_{\mu\nu}$ was
  identified with the 'Cotton tensor' because it reduces to its
  $3$-dimensional counterpart in certain symmetric cases. However, a
  higher-dimensional Cotton tensor already
  exists~\cite{Garcia:2003bw}, and its definition is different from
  that of Eq.~(\ref{C-def}). To avoid confusion, we refer to $C^{}_{\mu\nu}$
  as the C-tensor instead of the Cotton tensor.}.
For this reason, the quantity $v^{}_{\mu} := \theta^{}_{;\mu}$ is
sometimes referred to as the embedding coordinate, since it embeds the
$3$-dimensional CS theory into a $4$-dimensional spacetime.

Diffeomorphism invariance is preserved provided an additional equation
is satisfied. This equation can be obtained by computing the covariant
divergence of the modified field equations
\be
\label{add-const}
C_{\mu \nu;}{}^{\mu} = \frac{1}{8} v_{\nu} \pont 
= 8 \pi T_{\mu \nu}^{{\rm{mat}};\mu}
\ee
In vacuum, $T_{\mu \nu}^{\rm{mat}} = 0$ and thus we are left with the
so-called Pontryagin constraint 
\be
\pont = 0\,, \label{pontryagin}
\ee
which is an additional equation that the metric tensor has to satisfy.

%%%%%%%%%%%%%%%%%%%%%%%%%%%%%%%%%%%%%%%%%%%%%%%%%%%%%%%%%%%%%%%%%%%%%%%%%%%%%%
\section{Birkhoff's Theorem in Chern-Simons Modified Gravity}
\label{birk}

Birkhoff's theorem states that the most general, spherically symmetric
solution to the vacuum Einstein equations is the Schwarzschild metric.
In CS modified gravity, this is not necessarily the case because the
C-tensor modifies the field equations. In~\cite{jackiw:2003:cmo} it
was shown that the Schwarzschild solution persists in CS modified
gravity, but this does not necessarily imply that Birkhoff's theorem
holds. Let us then study Birkhoff's theorem in CS modified gravity in
more detail.

The line element of a general, spherically-symmetric spacetime can be
written as a warped product of two $2$-dimensional 
metrics~\cite{Gerlach:1979rw,Gerlach:1980tx}: a
Lorentzian one, $\met^{}_{AB}$ ($A,B,\ldots,H = t,r$), and the unit 
two-sphere metric, $\Omega^{}_{ab}$ ($a,b,\ldots,h = \theta,\phi$).  
This line element takes the following $2+2$ form:
\begin{equation}
g^{}_{\mu\nu}dx^\mu dx^\nu = \met^{}_{AB}(x^C)dx^A dx^B 
+r^2(x^A)\Omega^{}_{ab}(x^c)dx^a dx^b\,,
\label{2p2metric}
\end{equation}
where the warped factor is the square of the area radial coordinate $r$.  
Covariant differentiation on the Lorentzian manifold is denoted by a bar
while on the $2$-sphere is denoted by a colon.  The case of the Schwarzschild
metric in Schwarzschild coordinates in given by:
\be
\met^{}_{AB}dx^A dx^B = -f dt^2 +f^{-1}dr^2\,,~~~
f = 1-\frac{2M}{r}\,, \label{schmetricsch}
\ee
where $M$ is the BH mass.

The Cotton tensor associated with this metric vanishes identically if
the scalar field has the following form~\footnote{A similar result is
  found in~\cite{Grumiller:2007rv}}:
\be
\label{gen-CS}
\theta=\bar\theta(x^A) + r(x^A)\,\Theta(x^a)\,.
\ee
This functional form is invariant under coordinate changes that leave
the $2+2$ structure of the metric [Eq.~(\ref{2p2metric})] invariant,
i.e. coordinate transformations: $x^A~\rightarrow~\tilde{x}^A=
\tilde{x}^A(x^B)\,,$ $x^a~\rightarrow~\tilde{x}^a=\tilde{x}^a(x^b)\,,$
and transformations on the unit two-sphere.  Moreover, one can show
that the two pieces of the C-tensor shown in Eqs.~(\ref{cspart1})
and~(\ref{cspart2}) vanish independently for the family of scalar
fields given in Eq.~(\ref{gen-CS}).  For most of this paper, we will
consider the particular case in which the scalar field also respects
the spherical symmetry of the background:
\be
\label{choice-theta}
\theta = \bar\theta(x^A)\,.
\ee

Let us comment some more on this result and place it in context. The
CS scalar field $\theta$ is usually assumed to depend only on {\em
  cosmic} time, $\theta = \theta(t)$. This assumption has its roots in
the work of Jackiw and Pi~\cite{jackiw:2003:cmo}, who further imposed
that $\dot{\theta}=d\theta/dt$ be constant, such that 
time-translation symmetry and space-time reparameterization of the spatial 
variables be preserved. With these assumptions, they showed that CS gravity 
can be interpreted as a $4$-dimensional generalization of $3$-dimensional
Cotton gravity, and that the Schwarzschild and
Friedman-Robertson-Walker solutions persist in the modified theory.
Later on, Smith, {\emph{et.~al.~}}~\cite{Smith:2007jm} argued that the
CS scalar field could represent some quintessence field that enforces
the arrow of time associated with cosmic expansion. 
In this paper we have considered the most general spherically-symmetric
spacetime and we have written its line element in the $2+2$ form that
follows from its warped geometric structure.  This form of the metric
is invariant under general coordinate transformations in the two 
$2$-dimensional manifolds, associated with this $2+2$ structure.
Therefore, it is not too surprising that if Birkhoff's theorem is
satisfied for the canonical choice of CS scalar field, it is also
satisfied for fields that depend arbitrarily on $t$ and $r$.

%%%%%%%%%%%%%%%%%%%%%%%%%%%%%%%%%%%%%%%%%%%%%%%%%%%%%%%%%%%%%%%%%%%%%%%%%%%%%%
\section{Perturbations of a Schwarzschild Black Hole}
\label{BHPT}

%----------------------------------------------------------------------------
\subsection{Basics}

We begin with a short summary of the basics of metric perturbations
about a Schwarzschild background.  The spacetime metric of a perturbed
BH can be written in the form
%&f
%
\be
\label{BHPT-split}
\met^{}_{\mu\nu} = \bar{\met}^{}_{\mu\nu} + h^{}_{\mu\nu}\,,
\ee
where $\bar{\met}^{}_{\mu\nu}$ is the background Schwarzschild metric
(geometric objects associated with it will have an overbar)
and $h^{}_{\mu\nu}$ is a generic metric perturbation.  Thanks to the
spherical symmetry of the background, we can expand the metric
perturbations in (tensor) spherical harmonics.  In this way, we can
separate the angular dependence in the perturbative equations, which
yields a much more simple system of equations: a system of 1+1 partial
differential equations (PDEs) in time and in the radial area
coordinate $r$.  In addition, we can distinguish between harmonics
with polar and axial parity~\footnote{Polar and axial modes as those which,
under a parity transformation, acquire a factor of $(-1)^l$ and $(-1)^{l+1}$
respectively.}, also called even and odd parity modes respectively.  
In GR, the perturbative field equations decouple
for modes of different parity, but this may not be the case for
alternative theories. 

Let us then split the metric perturbation $h^{}_{\mu\nu}$ into polar and
axial perturbations, $h_{\mu\nu} = h^{\mbox{\small a}}_{\mu\nu} +
h^{\mbox{\small p}}_{\mu\nu}$, and each of these into (tensor) spherical
harmonics via
\begin{equation}
h^{\mbox{\small a}}_{\mu\nu} =
\sum_{\ell,m} h^{\mbox{\small a},\ell m}_{\mu\nu}\,,~~~
h^{\mbox{\small p}}_{\mu\nu} =
\sum_{\ell,m} h^{\mbox{\small p},\ell m}_{\mu\nu}\,,
\label{metricperturbations1}
\end{equation}
where
\begin{eqnarray}
h^{\mbox{\small a},\ell m}_{\mu\nu} = \left( \begin{array}{cc}
  0  &  h_A^{\ell m} \, S^{\ell m}_a \\
\\
 \ast  &  \  H^{\ell m} \, S_{ab}^{\ell m}
 \end{array}\right)\,, 
\label{maxial}
\end{eqnarray}
\begin{eqnarray}
h^{\mbox{\small p},\ell m}_{\mu\nu} = \left( \begin{array}{cc}
  h_{AB}^{\ell m}\, Y^{\ell m} &  p_A^{\ell m} \, Y^{\ell m}_a \\
\\
\ast  & \  r^2(K^{\ell m} \, Y^{\ell m}_{ab} +  G^{\ell m} \, Z_{ab}^{\ell m})
\end{array}\right)\,,
\label{mpolar}
\end{eqnarray}
and where asterisks denote components given by symmetry.  The quantity
$Y^{\ell m}$ refers to the standard scalar spherical harmonics
[see~\cite{Sopuerta:2006wj} for conventions], while $Y^{\ell m}_a$ and
$S^{\ell m}_a$ are polar and axial, vector spherical harmonics,
defined only for $\ell \ge 1$ via
\begin{equation}
Y^{\ell m}_a\equiv Y^{\ell m}_{:a}\,, \qquad
S^{\ell m}_a\equiv \eta^{}_a{}^b\,Y^{\ell m}_b\,.
\label{vec-sph-harm}
\end{equation}
Similarly, $Y^{\ell m}_{ab}$ and $Z_{ab}^{\ell m}$ are polar, and
$S_{ab}^{\ell m}$ axial, symmetric tensor spherical harmonics, defined
only for $\ell \ge 2$ via
\begin{eqnarray}
Y_{ab}^{\ell m} \equiv Y^{\ell m}\Omega^{}_{ab}\,,~~
Z^{\ell m}_{ab} \equiv Y^{\ell m}_{:ab}+\frac{\ell(\ell+1)}{2}Y^{\ell m}
\Omega_{ab}\,,  \label{ten-sph-harm1} 
\end{eqnarray}
\begin{eqnarray}
S^{\ell m}_{ab} \equiv S^{\ell m}_{(a:b)}\,.  \label{ten-sph-harm2}
\end{eqnarray}
The sign convention for the Levi-Civita tensor of the $2$-sphere is:
$\eta^{}_{\theta\varphi} = \sin{\theta}$.  All metric
perturbations, scalar ($h_{AB}^{\ell m}$), vectorial ($p_A^{\ell m}$
and $q_A^{\ell m}$), and tensorial ($K^{\ell m}\,$, $G^{\ell m}\,$,
and $q_2^{\ell m}$), are functions of $t$ and $r$ only.

In GR, the Einstein equations can be decoupled in terms of complex
master functions that obey wave-like master equations.  Once the
master functions are constructed we can recover all remaining metric
perturbations from them.  For axial modes, we can use the
Cunningham-Price-Moncrief (CPM) master
function~\cite{Cunningham:1978cp}, defined by
\begin{equation}
\Psi^{\ell m}_{\CPM} = -\frac{r}{\lambda^{}_\ell}\left( 
h^{\ell m}_{r,t} -  h^{\ell m}_{t,r} + \frac{2}{r}h^{\ell m}_t \right)\,, 
\label{PsiCPM_sch}
\end{equation}
whereas for polar modes we can use the Zerilli-Moncrief (ZM) master
function~\cite{Zerilli:1970fj,Moncrief:1974vm}
\begin{eqnarray}
\Psi^{\ell m}_{\ZM} &=& \frac{r}{1+\lambda^{}_\ell}\left\{ K^{\ell m}
  + (1+\lambda^{}_\ell)G^{\ell m}
  \right. 
\nonumber \\
& + & \left. \frac{f}{\Lambda^{}_\ell}\left[ f h^{\ell
  m}_{rr}-r  K^{\ell m}_{,r} -
  \frac{2}{r}(1+\lambda^{}_\ell)p^{\ell m}_r \right] \right\} \,, 
\label{PsiZM_sch}
\end{eqnarray}
where $\lambda^{}_\ell = (\ell+2)(\ell-1)/2$ and $\Lambda^{}_\ell =
\lambda^{}_\ell + 3M/r$.  
These two complex master
functions are gauge invariant.  In order to simplify our analysis
we shall here fix the gauge by setting the following metric
perturbations to zero (Regge-Wheeler gauge):
\ba
& & H^{\ell m} = 0\,, \\
& & G^{\ell m} = p_t^{\ell m} = p_r^{\ell m} = 0\,.
\ea
The master equations for the master functions, in the case of
perturbations without matter sources, have the following wave-like
structure:
\begin{equation}
\left[-\partial^{2}_{t^2} + \partial^{2}_{r^{2}_{\star}}
 - V^{\POLAR/\AXIAL}_\ell(r)\right]\Psi_{\CPM/\ZM}^{\ell m} = 0 \,,
\label{masterequations} 
\end{equation}
where $r^{}_{\star}$ is the {\em tortoise} coordinate $r^{}_{\star} = r +
2M\ln\left[r/(2M)-1\right]$.  The quantity $V^{\POLAR/\AXIAL}_\ell(r)$ is
a potential that depends on the parity and harmonic
number $\ell$ (its precise form can be found elsewhere.  To follow our
notation, see~\cite{Sopuerta:2006wj}). 

When the master functions are known, one can use them to construct
the plus and cross-polarized gravitational waveforms via
\begin{equation}
h^{}_{+} - i h^{}_{\times} = \frac{1}{2r}\sum^{}_{\ell\geq 2, m}
\sqrt{\frac{(\ell+2)!}{(\ell-2)!}}\left( \Psi_{\ZM}^{\ell m}+ i \Psi_{\CPM}^{\ell m}
\right) {}^{}_{-2}Y^{\ell m} \,, \label{waveforms}
\end{equation}
where ${}^{}_{-2}Y^{\ell m}$ denotes spherical harmonics of spin
weight $-2$~\cite{Goldberg:1967sp}.  One can also compute the fluxes
of energy and angular momentum emitted toward infinity and also into
the BH horizon in terms of the master functions.  These fluxes are
evaluated using a short-wavelength approximation in the {\em radiation
  zone}, where we can introduce a well-defined gauge-invariant
energy-momentum tensor for gravitational
radiation~\cite{Isaacson:1968ra,Isaacson:1968gw,Misner:1973cw}.
However, the expression for this effective energy-momentum tensor
depends on the structure of the field equations.  In the case of
CS modified gravity, given that the modification in the field
equations is additive, the form of these fluxes will be the same as in
GR but with extra terms proportional to the CS scalar field and its
derivatives.  That is,
\begin{eqnarray}
\dot{E}^{}_{\GW} & =  &\frac{1}{64\pi}\sum^{}_{\ell\geq 2, m}
\frac{(\ell+2)!}{(\ell-2)!}\left(
|\dot{\Psi}_{\CPM}^{\ell m}|^2 + |\dot{\Psi}_{\ZM}^{\ell m}|^2\right) 
\nonumber \\
& + & {\cal O}(\partial\theta,\partial^2\theta) \,,
\label{gwenergyflux}
\end{eqnarray}
\begin{eqnarray}
\dot{L}^{}_{\GW} & = & \frac{1}{64\pi}\sum^{}_{\ell\geq 2, m}i m
\frac{(\ell+2)!}{(\ell-2)!}\left(
\bar{\Psi}_{\CPM}^{\ell m}\dot{\Psi}_{\CPM}^{\ell m} +
\bar{\Psi}_{\ZM}^{\ell m}\dot{\Psi}_{\ZM}^{\ell m}\right)
\nonumber \\
& + & {\cal O}(\partial\theta,\partial^2\theta) \,,
\label{gwangularmomentumflux}
\end{eqnarray}
where the dots here denote time differentiation.

%----------------------------------------------------------------------------
\subsection{Pontryagin Constraint}\label{pont}

We have seen that diffeomorphism invariance requires an extra
condition, the Pontryagin constraint
[Eqs.~(\ref{CSmodification})-(\ref{pontryagin})].  This condition is
automatically satisfied not only by the most general spherically
symmetric metric (its C-tensor vanishes), but also for linear
perturbations of Minkowski spacetime. Nonetheless, this condition is
not satisfied for generic perturbations of a Schwarzschild BH.  In
this case, at first-order, the Pontryagin constraint becomes
\ba
\pont = \frac{96 M}{r^6} \left[h_t^{\ell m} + \frac{r}{2} \left(
 h_{r,t}^{\ell m} -  h_{t,r}^{\ell m} \right) \right]
\ell \left( \ell +1 \right) Y^{\ell m}\,.  
\label{pconsch}
\ea
Remarkably, the Pontryagin constraint involves only axial modes, which
is a consequence of the appearance of the Levi-Civita tensor (a
completely antisymmetric tensor) in the modification of the
gravitational sector of the action.  Perhaps even more remarkably,
this specific combination of axial modes corresponds exactly to the
Cunningham-Price-Moncrief master function~\cite{Cunningham:1978cp}
[eq.~(\ref{PsiCPM_sch})], which appears in the metric
waveforms~[Eq.~\eqref{waveforms}] and in the fluxes of energy and
angular momentum~[Eq.~\eqref{gwenergyflux}
and~\eqref{gwangularmomentumflux}].  We can thus rewrite
Eq.~(\ref{pconsch}) as
\be
\pont = - \frac{24 M}{r^6} \frac{\left(\ell + 2 \right)!}{\left(\ell -
    2 \right)!} \Psi_{\CPM}^{\ell m} Y^{\ell m}\,. 
\label{deltapont}
\ee
Then, the Pontryagin constraint forces the CPM function to vanish for
all harmonics with $\ell\geq 2$.  Such a restriction does not imply
that all axial perturbations necessarily vanish in CS modified
gravity, but it does require that these modes satisfy the relation
\be
\label{pont2}
h_{r,t}^{\ell m} =  h_{t,r}^{\ell m} - \frac{2}{r} h_{t}^{\ell
  m}\,. 
\ee
Such a condition seems to reduce the set of possible solutions of the
perturbative vacuum field equations, which might lead to an
overconstrained system. For the choice of $\theta$ in
Eq.~(\ref{choice-theta}), we shall see in Sec.~\ref{one-handed} that
this is indeed the case.

%----------------------------------------------------------------------------
\subsection{Structure of the Modified Field Equations}
\label{structure}

The CS modified dynamics of linear perturbations about a Schwarzschild
background can be studied by expanding the metric perturbations into
spherical harmonics, Eqs.~(\ref{metricperturbations1})-(\ref{mpolar}). 
After introducing these expansions into the field equations
\be
\label{mod-fiel-eq}
G^{}_{\mu\nu} = - C^{}_{\mu\nu}\,.
\ee
we can extract individual evolution equations for each harmonic.

The choice of the CS scalar field will determine the form of the
right-hand side in Eq.~\eqref{mod-fiel-eq}. Since we are considering
perturbations about a Schwarzschild spacetime, we must choose $\theta$
such that Birkhoff's theorem holds. In Sec.~\ref{birk}, we determined
that scalar fields of the form of Eq.~\eqref{gen-CS} would indeed
allow Birkhoff's theorem to persist in CS modified gravity. We shall
here choose $\theta$ as in Eq.~\eqref{choice-theta}, such that this
field agrees with the symmetries of the background, namely
\be
\label{gen-CS2}
\theta = \bar\theta(t,r)\,,
\ee
where $t$ and $r$ are here the Schwarzschild time and radial
coordinates. This family of CS coupling functions encompasses the
canonical choice: $\bar\theta = t/\mu$.

The structure of the linearized field equations is analyzed by first
looking at the harmonic decomposition of the Einstein tensor and the
C-tensor.  The structure of the former is well-known from the study of
perturbations of non-rotating BHs in GR, and is given by
\begin{eqnarray}
G^{\ell m}_{AB} & = & {\cal G}^{\ell m}_{AB}[\mb{U}^{\ell
  m}_{\POLAR}]\, Y^{\ell m}\,,  
\label{EinsteinABstruct} \\
G^{\ell m}_{Aa} & = & {\cal G}^{\ell m}_{A}[\mb{U}^{\ell m}_{\POLAR}]\, Y^{\ell m}_a
                    + {\cal H}^{\ell m}_{A}[\mb{U}^{\ell m}_{\AXIAL}]\, S^{\ell m}_a\,, 
\label{EinsteinAastruct}                    \\
G^{\ell m}_{ab} & = & {\cal G}^{\ell m}[\mb{U}^{\ell m}_{\POLAR}]\, Y^{\ell m}_{ab}
                    + {\cal H}^{\ell m}[\mb{U}^{\ell m}_{\POLAR}]\, Z^{\ell m}_{ab} \nn \\
                & + & {\cal I}^{\ell m}[\mb{U}^{\ell m}_{\AXIAL}]\, S^{\ell m}_{ab}\,,                    
\label{Einsteinabstruct}                
\end{eqnarray}
where $\mb{U}^{\ell m}_{\POLAR}$ denotes the set of $(\ell,m)$-polar
perturbations 
\begin{equation}
\mb{U}^{\ell m}_{\POLAR} = (h^{\ell m}_{AB}\,, p^{\ell m}_A\,,
K^{\ell m}, G^{\ell m}) \,,
\end{equation}
and $\mb{U}^{\ell m}_{\AXIAL}$ denotes the set of $(\ell,m)$-axial
perturbations 
\begin{equation}
\mb{U}^{\ell m}_{\AXIAL} = (h^{\ell m}_A\,, H^{\ell m} )\,.
\end{equation}
Expressions for the coefficients ${\cal G}^{\ell m}_{AB}$, ${\cal
  G}^{\ell m}_{A}$, ${\cal G}^{\ell m}$, ${\cal H}^{\ell m}_{A}$,
${\cal H}^{\ell m}$, and ${\cal I}^{\ell m}$ are given in
Appendix~\ref{explicitexpressions}.  Clearly, polar spherical
harmonics have functional coefficients that depend only on polar
metric perturbations, while axial spherical harmonics have functional
coefficients that depend only on axial metric perturbations.  On the
other hand, the harmonic structure of the C-tensor is given by
\begin{eqnarray}
C^{\ell m}_{AB} & = & {\cal C}^{\ell m}_{AB}[\mb{U}^{\ell 
                    m}_{\AXIAL}]\, Y^{\ell m}\,,  
\label{CottonABstruct} \\
C^{\ell m}_{Aa} & = & {\cal C}^{\ell m}_{A}[\mb{U}^{\ell
                    m}_{\AXIAL}]\, Y^{\ell m}_a
                    + {\cal D}^{\ell m}_{A}[\mb{U}^{\ell
                    m}_{\POLAR}]\, S^{\ell m}_a\,,  
\label{CottonAastruct} \\
C^{\ell m}_{ab} & = & {\cal C}^{\ell m}[\mb{U}^{\ell m}_{\AXIAL}]\,
                    Y^{\ell m}_{ab}
                    + {\cal D}^{\ell m}[\mb{U}^{\ell m}_{\AXIAL}]\,
                    Z^{\ell m}_{ab} \nn \\ 
                & + & {\cal E}^{\ell m}[\mb{U}^{\ell m}_{\POLAR}]
                    S^{\ell m}_{ab}\,,    
\label{Cottonabstruct}                
\end{eqnarray}
where explicit expressions for the coefficients ${\cal C}^{\ell
  m}_{AB}$, ${\cal C}^{\ell m}_{A}$, ${\cal C}^{\ell m}$, ${\cal
  D}^{\ell m}_{A}$, ${\cal D}^{\ell m}$, and ${\cal E}^{\ell m}$ are
also given in Appendix~\ref{explicitexpressions}.  In this case, polar
spherical harmonics have functional coefficients that depend on axial
perturbations, while axial spherical harmonics have functional
coefficients that depend on polar perturbations.  The main consequence
of this fact is that in CS modified gravity modes with different
parity are coupled, and hence, in general they cannot be treated
separately.

The linearized field equations, after harmonic decomposition, become
\ba
\label{decoup1}
{\cal{G}}^{\ell m}_{AB} &=& - {\cal{C}}^{\ell m}_{AB}\,, 
\qquad
{\cal{G}}^{\ell m}_{A} = -  {\cal{C}}^{\ell m}_{A}\,, 
\\
{\cal{G}}^{\ell m} &=& -  {\cal{C}}^{\ell m}\,, 
\qquad
{\cal{H}}^{\ell m} = -  {\cal{D}}^{\ell m}\,,
\label{decoup1bis}
\\
{\cal{H}}^{\ell m}_A &=& - {\cal{D}}^{\ell m}_A\,,
\qquad
{\cal{I}}^{\ell m} = -  {\cal{E}}^{\ell m}\,.
\label{decoup2}
\ea
We can view these equations as the standard Einstein equations,
linearized about a Schwarzschild background, with ``source terms''
that depend linearly on metric perturbations of opposite parity and
their derivatives.  Such a coupling between different parity modes is
analogous to what occurs to gravitational-wave perturbations about a
Minkowski background, where left- and right-polarized perturbations
mix~\cite{jackiw:2003:cmo,alexander:2004:lfg,alexander:2005:bgw}.  The
intrinsic decoupling of the modified field equations into polar and
axial sectors is thus lost.

%%%%%%%%%%%%%%%%%%%%%%%%%%%%%%%%%%%%%%%%%%%%%%%%%%%%%%%%%%%%%%%%%%%%%
\section{Black Hole Perturbation Theory with a Canonical Embedding}
\label{BHPT-canon}

In order to understand the dynamics of the perturbations that derives
from this theory and, in particular, the role and consequences of the
Pontryagin constraint, we shall concentrate on the special case of a
canonical CS coupling function, {\emph{i.e.}}~$\bar\theta = t/\mu$.
This canonical coupling function leads to the canonical timelike
embedding $v^{}_{\mu} = [1/\mu,0,0,0]$ and its acceleration
$v^{}_{\mu;\nu} = \Gamma_{\mu\nu}^{\sigma} v^{}_{\sigma}$.

%%%%%%%%%%%%%%%%%%%%%%%%%%%%%%%%%%%%%%%%%%%%%%%%%%%%%%%%%%%%%%%%%%%%%%%%%%%
\subsection{One-handed Perturbations}\label{one-handed}

To begin with, we concentrate on perturbations with a single handedness,
{\emph{i.e.}}~purely polar or purely axial perturbations. We shall
analyze these cases separately.

%----------------------------------------------------------------
\subsubsection{Pure Axial Perturbations}

We begin with pure axial metric perturbations, that is
\be
h_{AB}^{\ell m} = p_A^{\ell m} = K^{\ell m} = G^{\ell m} = 0 \,.
\ee 
This conditions imply 
\ba
& & {\cal{G}}^{\ell m}_{AB} = {\cal{G}}_A^{\ell m} = 
{\cal{G}}^{\ell m} = {\cal{H}}^{\ell m} = 0\,, \\
& & {\cal{D}}^{\ell m}_A = {\cal{E}}^{\ell m} = 0\,, 
\ea
and hence the field equations reduce to:
\ba
\label{pure-axial-einst}
{\cal{H}}^{\ell m}_A = 0\,, & \qquad & {\cal{I}}^{\ell m} = 0\,, \\
{\cal{C}}^{\ell m}_{AB} = 0\,, & \qquad &
{\cal{C}}^{\ell m}_A = 0\,, \label{pure-axial-CS} \\
{\cal{C}}^{\ell m} = 0\,, & \qquad &
{\cal{D}}^{\ell m} = 0\,. \label{pure-axial-CS2}
\ea

Looking at the expressions of ${\cal{C}}^{\ell m}$ and
${\cal{C}}^{\ell m}_{rr}$ (see Appendix~\ref{explicitexpressions}) we
see that both of them are proportional to the metric perturbation
$h_{r}^{\ell m}$.  Therefore, using
Eqs.~(\ref{pure-axial-CS})-(\ref{pure-axial-CS2}) we conclude that 
\be
h_{r}^{\ell m} = 0\,.
\ee
Similarly, ${\cal{C}}^{\ell m}_{tr}$ is proportional to $h_t^{\ell
  m}$, and thus, using Eq.~\eqref{pure-axial-CS2} 
\be
h_{t}^{\ell m} = 0\,.
\ee
Since $H^{\ell m}$ is zero in the Regge-Wheeler gauge, we conclude
that all axial perturbations must vanish.  In summary, the Pontryagin
constraint together with the coupling of opposite parity modes in the
modified field equations, forbids the existence of purely polar
oscillations of a Schwarzschild BH in CS modified gravity with a
canonical embedding.

%-------------------------------------------------------
\subsubsection{Pure Polar Perturbations}
Let us now study pure polar perturbations by setting all axial modes
to zero:
%&i
\be 
h^{\ell m}_A = 0\,. 
\ee
The immediate consequences are
\ba
& & {\cal{H}}^{\ell m}_A = {\cal{I}}^{\ell m} = 0\,,\\ 
& & {\cal{C}}^{\ell m}_{AB} = {\cal{C}}^{\ell m}_A = 
    {\cal{C}}^{\ell m} = {\cal{D}}^{\ell m} = 0\,,
\ea
and the field equations become
\ba
\label{pure-polar-einst}
& &{\cal{G}}_{AB}^{\ell m} = {\cal{G}}_A^{\ell m} = {\cal{G}}^{\ell m}
= {\cal{H}}^{\ell m}= 0\,, \\
\label{pure-polar-CS}
& &{\cal{D}}_A^{\ell m} = {\cal{E}}^{\ell m} = 0\,. 
\ea
From ${\cal{H}}^{\ell m} = 0$ we find that (see
Appendix~\ref{explicitexpressions})
\be
h_{tt}^{\ell m} = f^2 h_{rr}^{\ell m}\,, 
\label{inter1} 
\ee
while, from equations ${\cal{G}}_{A}^{\ell m} = 0 = {\cal{G}}^{\ell
  m}_{AB}$ we find expressions for all first and second derivatives of
$K^{\ell m}$ in terms of $h_{rr}^{\ell m}$,
$h_{tr}^{\ell m}$ and its derivatives.  Substituting these expressions
into ${\cal{E}}^{\ell m} = 0$ leads to
\be
h_{rr}^{\ell m} = 0\,, 
\ee
which combined with Eq.~(\ref{inter1}) implies
\be
h_{tt}^{\ell m} =0\,.
\ee
There are now only two non-zero metric perturbations: $K^{\ell m}$ and
$h_{tr}^{\ell m}$.  Inserting the expressions for the derivatives of
$K^{\ell m}$ into ${\cal{D}}_A^{\ell m}=0$, we can solve the resulting
equations for $ h_{tr,rr}^{\ell m}$ and $ h_{tr,tr}^{\ell m}$.  Using
all the information we have collected so far in ${\cal{G}}^{\ell m} =
0$, one can show by direct evaluation that this equation requires that
$ h_{tr,t}^{\ell m} = 0$.  Such a result, in combination with the
previously found expression for $ h_{tr,tr}^{\ell m}$, leads to 
\be
K^{\ell m} = 0\,. 
\ee
Returning to the previous expressions for derivatives of $K^{\ell m}$
we obtain
\be
h_{tr}^{\ell m} = 0\,.
\ee
We have then found that all polar perturbations vanish. We conclude
that purely polar oscillations of a Schwarzschild BH are not allowed
in CS modified gravity with a canonical embedding.

The results obtained for single-parity oscillations are quite
surprising and raise questions about its robustness. In other words,
can we expect the same conclusion if we repeat the analysis for other
CS coupling functions?  We have repeated this analysis for different
CS coupling functions within the class $\theta=\bar\theta(t,r)$.  The
algebra involved is significantly more complicated and it requires
intensive use of symbolic manipulation software~\cite{grtensor}. In
all studied cases, we have arrived at the same conclusion: CS modified
gravity does not allow for single-parity BH oscillations.

%----------------------------------------------------------------
\subsubsection{General Perturbations}

General perturbations are significantly more difficult to analyze,
since we cannot separately study the components of the Einstein and
C-tensors, as was the case for single parity oscillations, due to the
particular structure of these tensors
[Eqs.~(\ref{decoup1})-(\ref{decoup2})].  Nonetheless, we can still
make some general comments about the consequences of the Pontryagin
constraint on general oscillations and the extent to which these are
restricted.

The Pontryagin constraint unavoidingly adds an extra equation to the
modified field equations, {\emph{i.e.}}~Eq.~(\ref{pont2}).  Since this
equation only involves axial perturbations, one may {\emph{a priori}}
think that polar modes are unaffected. The modified field equations,
however, mix polar and axial modes, and thus, polar modes are also
affected and restricted by the Pontryagin constraint.  This
situation has been clearly illustrated by the study of single-parity
oscillations in the previous subsection.

The coupling of polar and axial metric perturbations in the modified
field equations, together with the extra condition imposed by the
Pontryagin constraint, lead to new conditions on the metric
perturbations. In other words, these new conditions constitute new
equations for the metric perturbations that are independent of the
previous ones.  For example, the components of ${\cal C}^{\ell
  m}_{AB}$ in Eq.~(\ref{CottonABstruct}) are linear combinations of
the axial metric perturbations $h^{\ell m}_A$ (see
Appendix~\ref{explicitexpressions}), which can be combined to
reconstruct the CPM master function, leading to an additional equation
for polar modes. Similarly, the components of ${\cal H}^{\ell m}_A$
[Eq.~(\ref{EinsteinAastruct})] can also be combined to construct the
CPM master function, and hence find another constraint on polar modes.
Again, due to the mixing of different parity modes, these constraints
on polar modes can in turn produce new constraints on axial modes. It
is unclear where this chain of new constraints on the metric
perturbations ends, but it is very likely that they will severely
restrict the set of allowed BH oscillations.

One may think that choosing the CS coupling function $\theta$
appropriately, namely choosing $\bar\theta(x^A)$ and $\Theta(x^a)$ in
Eq.~(\ref{gen-CS}), may end this chain of new constraints through
cancellations in the generation of new equations. In the case
$\theta=\bar\theta(x^A)$, if such cancellations were to occur for a
certain harmonic number $\ell$, they could not happen for other
$\ell$, as the equations depend on the harmonic number.  Moreover,
such cancellations would have to occur in different ways, associated
with the different combinations of equations that produce new
constraints, discussed in the previous paragraph. Adding the term
$\Theta(x^a)$ to the CS scalar field only complicates the field
equations further by inducing a coupling between perturbations with
different harmonic number. Even in this case, given that the
functional coefficient of $\Theta(x^a)$ is fixed (it is just $r$) and
$\ell$-independent, it seems unlikely that the coupling of harmonic
modes will lead to the cancellations necessary to avoid new equations
for the metric perturbations.

Our analysis suggests that the present structure of CS modified
gravity does not seem to allow for generic BH oscillations.  In
particular, we have shown that for certain choices of the CS scalar
field, single-parity perturbations are not allowed.  For general
perturbations, and with the help of computer algebra, we have analyzed
the equations as discussed above and found that cancellations in the
generation of new conditions on the metric perturbations are very
unlikely.  Therefore, the present set up for CS modified gravity does
not seem to allow for the study of generic oscillations of
non-rotating BHs. Such a result is reminiscent to that found in GR
when additional restrictions are imposed on the metric tensor.  An
example of this can be found in the study of relativistic cosmological
dynamics~\cite{Sopuerta:1997nh,Sopuerta:1998rt}

%The modified theory seems too restrictive, since the constraints that
%it imposes on the dynamics of BH oscillations seem to eliminate all
%possible perturbations. Nonetheless, such constraints could be avoided
%by slightly extending the modified theory, while keeping its main
%characteristics untouched. Even within the non-extended modified theory, we can still
%discuss some of its physical consequences.  Two distinctive features
%of CS modified gravity are the coupling of different parity modes and
%the Pontryagin constraint.  While the former can be preserved by
%extensions of the modified theory, the latter must be relaxed to
%prevent the vanishing of the CPM master function.  Since, by
%assumption, any extension should introduce small modifications to CS
%gravity, the CPM master function would be forced to be small but not
%quite zero.  If this is the case, such extensions of the modified
%theory would have strong implications in the energy flux and, in
%particular, in the angular momentum flux of gravitational
%waves~[Eqs.~\eqref{gwenergyflux} and~\eqref{gwangularmomentumflux}].
%These arguments motivate the study of BH oscillations in extensions of
%CS modified gravity, which we shall explore in the next section.

Although the modified theory seems too restrictive, it is possible to
develop an extension where the aforementioned constraints are avoided,
while keeping the main characteristics of the modified theory
untouched. Two distinctive features of the non-extended theory are the
coupling of different parity modes and the Pontryagin constraint.
While the former can be preserved by extensions of the modified
theory, the latter must be relaxed to prevent the vanishing of the CPM
master function.  Since, by assumption, any extension should introduce
small modifications to CS gravity, the CPM master function would be
forced to be small but not quite zero.  If this is the case, such
extensions would have strong implications in the energy flux and, in
particular, in the angular momentum flux of gravitational
waves~[Eqs.~\eqref{gwenergyflux} and~\eqref{gwangularmomentumflux}].
These arguments motivate the study of BH oscillations in extensions of
CS modified gravity, which we shall explore in the next section.

%What would be some of the physical consequences if CS modified gravity
%did not allow for BH oscillations to linear order? Consider a small
%star falling into a supermassive BH. In GR, such an event would lead
%to quasinormal ringing of the BH, which would ``lose its hair'' by
%radiating energy and angular momentum in gravitational waves. If such
%oscillations are not allowed, the BH cannot radiate its hair and
%energy and angular momentum would never decrease. The latter effect
%could have radical consequences, since it would imply that the cosmic
%censorship conjecture is violated. This is so because after numerous
%encounters with stars there would a non-vanishing probability that the
%angular momentum of the BH would exceed the Kerr limit,
%{\emph{i.e.}}~$J_{\textrm{BH}} > M_{\textrm{BH}}^2$, where
%$J_{\textrm{BH}}$ and $M_{\textrm{BH}}$ are the spin angular momentum
%and mass of the BH respectively.

%%%%%%%%%%%%%%%%%%%%%%%%%%%%%%%%%%%%%%%%%%%%%%%%%%%%%%%%%%%%%%%%%%%%%%%%%%%%%%
\section{Beyond the Cannon}\label{beyond}

The strong restrictions obtained on the dynamics of BH oscillations
thus far can be bypassed provided we considered extensions of CS
modified gravity beyond the canon. One such possibility is to consider
a more general CS coupling function that, for example, is not spherically
symmetric.  However, as we have argued before, it does not seem likely
that such a modification would avoid the generation of new
restrictions on the metric perturbations.  Another more interesting
possibility is to extend the action of Eq.~\eqref{full-action} to
allow for the dynamical evolution of the CS scalar field. Such a route
is particularly promising because it weakens the Pontryagin
constraint, as we shall see in this section.  However, a certain
amount of arbitrariness is inherent to such a route, encoded in the
choice of the scalar field action and, in particular, its potential.
We shall discuss this and other issues in the remaining of this
section.

%----------------------------------------------------------------------
\subsection{Extended CS Modified Gravity}

Until now, we have treated the CS coupling function as a
non-dynamical quantity.  Recently, Smith, et.~al.~\cite{Smith:2007jm}
added a kinetic and a potential term to the action, which they found
did not contribute to their weak-field analysis. These additional
terms are of the form
\be
S^{}_{\rm{ext}} = S^{}_{\rm{CS}} + \kappa \int d^4x  \sqrt{-g} \left[ 
  \frac{1}{2} \theta_{,\mu} \theta^{,\mu} - V(\theta)
\right]\,,  
\ee
where $V(\theta)$ is some potential for the CS scalar field. 

The variation of the extended action with respect to the metric and
scalar field yields the equations of motion of the extended theory
\ba
\label{ext-mod-field}
G_{\mu \nu} + C_{\mu \nu}  &=& 8 \pi
\left(T_{\mu \nu}^{\rm{mat}} + T_{\mu \nu}^{\theta} \right)\,,
\\ 
\label{CS-EOM}
\square \theta &=& -\frac{dV}{d\theta}  - \frac{1}{4}\pont\,,
\ea
where the D'Alambertian of any scalar can be computed from 
\ba
\square \theta = \frac{1}{\sqrt{-g}} \left[\sqrt{-g} \; g^{\mu \nu} \; 
\theta_{,\nu}\right]_{,\mu}\,,
\ea
and where the stress-energy tensor of the scalar field
is~\cite{Carrol}
\be
T_{\mu \nu}^{\theta} = \theta_{,\mu}  \theta_{,\nu} -
\frac{1}{2} g_{\mu \nu} \theta_{;\sigma} \theta_{;}^{\sigma} - g_{\mu
  \nu} V(\theta)\,.  
\ee
In this extension of CS modified gravity, the Pontryagin constraint is
replaced by an equation of motion for the CS scalar field,
Eq.~\eqref{CS-EOM}, where the quadratic curvature scalar $\pont$ plays
the role of a driving force.  Moreover, looking at the field equations
for the metric, we realize that they are not only sourced by the
matter fields but also by the CS scalar field through its
energy-momentum content.  In the next subsection we shall study what
consequences this modification imposes on the satisfaction of
Birkhoff's theorem and the Pontryagin constraint.

%----------------------------------------------------------------------
\subsection{Birkhoff's Theorem in the Extended Theory}

The simple derivation of Birkhoff's theorem in Sec.~\ref{birk} is now
modified by the extended action. For a CS scalar field of the form of
Eq.~\eqref{gen-CS} and the line element of Eq.~\eqref{2p2metric}, the C-tensor 
vanishes, which now leads to
\be
G_{\mu \nu} = 8 \pi T_{\mu \nu}^{\theta}.
\ee
If the metric of Eq.~\eqref{2p2metric} is a solution to Einstein's
equations, as is the case for the Schwarzschild metric, the
stress-energy tensor of the scalar field must vanish.  For the choice
of $\theta$ in Eq.~(\ref{gen-CS2}) and the Schwarzschild metric, this
implies the following conditions: \ba
\label{Tcs}
0 &=& \theta_{,t} \theta_{,r},
\nonumber \\
0 &=& \frac{1}{2} \theta_{,t}^2 + \frac{f^2}{2} \theta_{,r}^2 \pm f
V(\theta),   
\ea
where $f$ is given in Eq.~(\ref{schmetricsch}). The only solution to
this system is the trivial one: $V = 0$ and $\theta = {\rm{const.}}$,
which reduces CS theory to GR. However, if we treat $\theta$ as a
small quantity, as it is suggested by the different physical scenarios
that motivate CS modified gravity, then Birkhoff's theorem holds to
${\cal{O}}(\theta^2)$ provided $V$ is at least of the same order.  An
example of such a potential is a mass term $V = m \theta^2$, typical
of scalar interactions.

The results of this subsection also apply to more general CS coupling
functions and background metrics. In fact, these results hold for any
line element that represents a general, spherically symmetric
spacetime, {\emph{i.e.}}~Eq.~\eqref{2p2metric}. Moreover, they also
hold for the most general CS field that leads to a vanishing C-tensor,
{\emph{i.e.}}~Eq.~\eqref{gen-CS}, because the kinetic sector of
$T_{\mu \nu}^{(\theta)}$ is always quadratic in $\theta$.

%----------------------------------------------------------------------
\subsection{BH perturbations in the Extended Theory and the Pontryagin
constraint}
\label{birk2}

In the extended theory, the Pontryagin constraint is replaced by a
dynamical equation for the CS scalar field with a purely gravitational
driving term.  Since now there is no Pontryagin constraint, there is
no {\em a priori} reason for the equations to disallow general BH
oscillations. In fact, as we shall see in this subsection, the
extended theory leads to a system of $11$ PDEs for $11$ dynamical
variables $(\theta,h_{\mu \nu})$.

But can we treat $\theta$ as a perturbation? As we have just seen,
Birkhoff's theorem holds only to linear order in $\theta$, for a wide
class of potentials $V(\theta)$. On the one hand, in most
string-theory scenarios that necessitate the CS
correction~\cite{Alexander:2004xd,Alexander:2007qe,Smith:2007jm}, the
scalar field $\theta$ is proportional to the string scale. In such
cases, $\theta$ is much smaller than any metric perturbation and
Birkhoff's theorem holds. Nonetheless, even within such frameworks,
the CS correction could be enhanced by couplings to non-perturbative,
string theoretical degrees of freedom ({\emph{i.e.}}~gravitational
instantons)~\cite{Svrcek:2006yi}. Moreover, there are some theoretical
frameworks where the string coupling $g_{\mathrm{s}}$ vanishes at late
times \cite{
Brandenberger:1988:sit,%
Tseytlin:1991:eos,%
Nayeri:2005:pas,%
sun:2006:ccm,%
wesley:2005:cct,%
Brandenberger:2001:lpi,%
battefeld:2006:sgc,%
brandenberger:2006:sgc,%
brandenberger:2007:sgc,%
brax:2004:bwc}, in which case a larger coupling is permissible. 

From this discussion, we cannot necessarily assume that the magnitude
of the scalar field is smaller or of the same order as the metric
perturbations. In fact, these perturbations and the scalar field act
on completely independent scales.  Therefore, all we can assume is
that $\theta$ and $|h_{ab}|$ are both {\emph{independently}} smaller
than the background, which is enough to justify the use of
perturbation theory.  

We shall thus consider a two-parameter (bivariate) perturbative
expansion of CS gravity. One perturbative parameter shall be
associated with the metric perturbations, $\epsilon$, and the other
with the scalar field, $\tau$. The metric in Eq.~\eqref{BHPT-split}
and the scalar field can then be rewritten as
\ba
\label{metricpert}
\met^{}_{\mu\nu} & = & \bar{\met}^{}_{\mu\nu} + \epsilon\,h^{}_{\mu\nu}\,, \\
\label{theta-chosen}
\theta & = & \tau(\bar{\theta} + \epsilon\,\delta\theta) \nonumber \\
       & = &  \tau\bar\theta + \tau\epsilon
\sum^{}_{\ell\geq 1, m} \tilde{\theta}^{\ell m} \, Y^{\ell m}\,,
\ea
where $\bar\theta$ satisfies Birkhoff's theorem [Eq.~(\ref{gen-CS})]
and respects the spherical symmetry of the background. The quantities
$\tilde\theta^{\ell m}=\tilde\theta^{\ell m}(x^A)$ are harmonic
coefficients of the scalar field perturbations, associated with the BH
oscillations.  There are no $\ell = 0$ modes in the sum of
Eq.~(\ref{theta-chosen}) because they can always be absorbed in the
monopole term $\bar\theta$.

The equations of motion for the metric perturbation and the scalar
field now become formal bivariate expansions in $\epsilon \ll 1$ as
well as $\tau \ll 1$. The modified field equations to zeroth order in
$\epsilon$ are simply equations for the background metric, which are
automatically satisfied to this order by Birkhoff's theorem.  The
equation of motion for the scalar field to the same order becomes
\be
\label{eqbartheta}
\tau \; \bar{\square}\theta = \tau \bar\met^{\mu\nu} \bar\nabla_{\mu}
\bar\nabla_{\nu} \theta^{}_{} = 0\,.
\ee
In Eq.~(\ref{eqbartheta}), and in all remaining equations, the
overhead bars on any quantity are to remind us that these quantities
are to be evaluated with respect to the background metric
$\bar{\met}_{\mu \nu}$. Moreover, in Eq.~\eqref{eqbartheta} and
henceforth, we shall neglect the contribution of the potential, by
assuming that it is at least of ${\cal{O}}(\tau^2)$. As discussed
earlier, the potential encodes a certain arbitrariness in the
extension of the CS modification, which is why we have chosen to
neglect it.

The first-order equations govern the dynamics of the metric
perturbation and the perturbations of the scalar field. The equation
of motion for the scalar field to ${\cal{O}}(\epsilon)$ is given by
\ba
\epsilon \frac{1}{4}\delta(\pont)
&=& \epsilon \tau\left\{\bar{\square}\delta\theta - \left[\bar\theta^{}_{,\mu\nu}
+ \left(\ln\sqrt{-\bar{g}}\right)^{}_{,\mu}\bar\theta^{}_{,\nu} \right]h^{\mu\nu}
\right. 
\nonumber \\
&-& \left. 
\bar\theta^{}_{,\mu}h^{\mu\nu}{}^{}_{,\nu} + \frac{1}{2} h_{,\mu}
\bar{\met}^{\mu \nu} \bar{\theta}_{,\nu} \right\} \,. \label{eqdeltatheta}
\ea
In Eq.~\eqref{eqdeltatheta}, $\delta(\pont)$ is the functional
coefficient of $\pont$ to ${\cal{O}}(\epsilon)$
[Eq.~(\ref{deltapont})] and $h = \met^{\mu \nu} h_{\mu \nu} =
\bar{\met}^{\mu \nu} h_{\mu \nu}$.  Equation~\eqref{eqdeltatheta} is
harmonically decomposed in the Regge-Wheeler gauge in
Eq.~(\ref{eqdeltathetarw}) of Appendix~\ref{rwexpressions}.

Similarly, the equations of motion for the metric perturbation to
${\cal{O}}(\epsilon)$ reduce to
\ba
{\cal O}(\tau^2) 
&=&
\epsilon \delta G^{}_{\mu\nu} + \epsilon \tau\left[
\bar\theta{}^{}_{,\sigma} \bar{\eta}^{\sigma\alpha
\beta}{}^{}_{(\mu} \bar\nabla_{\alpha} \delta R^{}_{\nu)\beta} 
\right. 
\nonumber \\
&+& \left.  
\left(\bar\nabla_{\tau} \bar{\theta}{}_{,\sigma}\right)
\delta{\,^\ast\!}R^\sigma{}_{(\mu}{}^{\tau}{}^{}_{\nu)} +
 \left(\bar\nabla_{\tau} \delta\theta^{}_{,\sigma}
\right. \right. 
\nonumber \\
&-& \left. \left. 
\bar{\theta}^{}_{,\rho} 
\delta\Gamma^{\rho}_{\sigma\tau}\right)\,
 {\,^\ast\!}\bar{R}^\sigma{}_{(\mu}{}^{\tau}{}^{}_{\nu)} \right]
\,. \label{eqdeltah} 
\ea
In Eq.~\eqref{eqdeltah}, $\delta G_{\mu \nu}$ is the functional
coefficient of $G_{\mu \nu}$ to ${\cal{O}}(\epsilon)$ (see
Appendix~\ref{explicitexpressions}), and all other $\delta A^{\alpha
  \beta \ldots}{}_{\chi \zeta \ldots}$ stands for the coefficient to
${\cal{O}}(\epsilon)$ of any tensor $A^{\alpha \beta \ldots}{}_{\chi
  \zeta \ldots}$. In fact, Eq.~\eqref{eqdeltah} is the formal
covariant expression of the perturbation of the Einstein and C-tensors
to ${\cal{O}}(\epsilon)$, which we computed in
Appendix~\ref{explicitexpressions} for the CS coupling function in
Eq.~\eqref{gen-CS2}, with the exception of the term proportional to
$\delta \theta$. The quadratic terms on the right-hand side of
Eq.~\eqref{eqdeltah} come from the energy-momentum tensor of the
scalar field $\theta$ [see Eq.~(\ref{ext-mod-field})].  When taking
the divergence of this equation, we recover Eq.~(\ref{eqdeltatheta})
only when quadratic terms in $\tau$ are taken into account because
functional differentiation with respect to $\theta$ reduces the order
in $\tau$ of the resulting expression by unity.

Equations~(\ref{eqdeltatheta}) and~(\ref{eqdeltah}) govern the
dynamics of BH oscillations in this extended version of CS modified
gravity, but how do we solve them?  Although it would be useful to
decouple these equations in terms of master functions as done in GR,
this is not an easy task. In addition to the mixing of parity modes
and the fact that the Pontryagin constraint can no longer be used to
simplify equations, there is now additional terms in the modified
field equations due to the perturbations of the scalar field.
Moreover, these perturbations possess their own dynamics, and hence,
any decoupling should involve the whole set of perturbative variables,
{\emph{i.e.}}~$(\tilde\theta^{\ell m},h^{\ell m}_{\mu \nu})$.

Nonetheless, it should be possible to numerically solve the
perturbative field equations of the extended modified theory in an
iterative way with $\tau \ll 1$.  One such possible iterative
procedure is as follows. To zeroth order in $\tau$, the equations
reduce to those of GR, which can be decoupled and solved numerically
with standard methods. The numerical result can then be reinserted in
the field equations to first order in $\tau$. These equations can now
be decoupled in exactly the same way and will contain source terms
determined by the zeroth order solution. This problem will be tackled
in a future publication.

Before concluding, let us make some general remarks about the
consequences of extending CS modified gravity through such kinetic
terms. Equation~(\ref{eqdeltatheta}) shows that the Pontryagin
constraint has become an evolution equation for the perturbations of
the scalar field, $\delta\theta$. Together with Eq.~\eqref{eqdeltah},
this evolution equation constitutes a system of PDEs for the
perturbative variables $(\tilde\theta^{\ell m},h^{\ell m}_{\mu \nu})$
that could in principle allow for generic oscillations.
%As is standard with evolution systems, such an evolution equation
%requires some initial conditions.  
However, since $\tau \ll 1$, the magnitude of the CPM master function
is forced to be small [of ${\cal{O}}(\tau)$].  Otherwise, the scalar
field would lead to an amplification of the CS correction to levels
forbidden by solar system tests~\cite{Smith:2007jm}. The extended
theory thus relaxes the vanishing of the CPM function and replaces it
by somewhat of a suppression of radiative axial modes.

Such an effect may lead to important observational consequences.
In particular, the dynamics of astrophysical systems, where axial
modes contribute significantly to the gravitational-wave emission,
would be greatly modified. In such cases, Eqs.~(\ref{gwenergyflux})
and~(\ref{gwangularmomentumflux}) suggest that the flux of energy
emitted would be dominated by polar modes and the flux of angular
momentum would be very small, since it would be linear in $\tau$. Such
a flux suppression is in contrast to predictions of GR, where the
gravitational wave emission of angular momentum is known to be large
(approximately $14 \%$ of the initial ADM angular momentum for
quasi-circular BH mergers~\cite{Campanelli:2005dd}).  Consequently,
the dynamics of gravitational wave sources in the radiation-reaction
dominated phase should be quite different in CS modified gravity
relative to GR, thus allowing for gravitational wave tests of the
extended theory.

%%%%%%%%%%%%%%%%%%%%%%%%%%%%%%%%%%%%%%%%%%%%%%%%%%%%%%%%%%%%%%%%%%%%%%%%%%%%%%
\section{Conclusions}
\label{conclusions}

We have studied Schwarzschild BH perturbation theory in CS modified
gravity. We began by showing that Birkhoff's theorem (the statement
that the Schwarzschild solution is the only vacuum,
spherically-symmetric solution of the theory) holds in the modified
theory for a wide class of CS coupling functions. We then decomposed
the metric perturbations into tensor spherical harmonics and found the
linearized modified field equations that determine their behavior.
The divergence of these equations led to the linearized Pontryagin
constraint, which imposes a restriction on axial metric perturbations
--the CPM master function has to vanish.

Once these preliminary issues were studied, we focused on the general
structure of the metric perturbations in CS modified gravity. We found
that the modified theory adds new terms to the field equations that
couple perturbations with polar and axial parity. Moreover, due to the
restrictions imposed by the Pontryagin constraint, we find that in
general the system of equations is overdetermined. Whether the entire
set of equations is compatible remains unclear, but for a wide class
of initial physical conditions, we found that linear BH oscillations
are not allowed in CS modified gravity. Specifically, we showed that
pure axial or pure polar oscillations are disallowed for a wide class
of coupling functions. 

Possible extensions of the modified theory that would allow for
generic BH oscillations were also discussed. In particular, we
investigated the possibility of providing the CS coupling function
with dynamics. In other words, the inclusion of a kinetic and
potential term in the action led to the replacement of the Pontryagin
constraint with an equation of motion for the coupling function. This
route then lifts the vanishing restriction of the Pontryagin
constraint and imposes a smallness condition on the CPM function.

The extended CS modified framework thus allows for generic BH
oscillations but it imposes an important smallness restriction on
axial perturbation that could lead to astrophysical observables. In
particular, we saw that such a suppression of axial modes would lead
to a significant decrease in the magnitude of energy, linear and
angular momentum carried by gravitational radiation relative to GR.
Such a decrease in gravitational wave intensity would have important
consequences in the dynamics of compact object inspirals, specially in
the radiation-reaction dominated phase. 

Future observations of the ringdown signal in binary BH mergers could
be used to test and constrain CS modified gravity. For such studies,
the results found in this paper would be critical in order to
determine the quasi-normal frequency spectrum of perturbations.
Future work could concentrate on such a spectrum, by numerically
studying the linearized and harmonically-decoupled field equations
presented here. Moreover, semi-analytic studies might also be possible
through the close-limit approximation. Only through detailed studies
of all aspects of the modified theory and its links to experimental
observations will we be able to determine its viability.

%%%%%%%%%%%%%%%%%%%%%%%%%%%%%%%%%%%%%%%%%%%%%%%%%%%%%%%%%%%%%%%%%%%%%%%%%%%%%%
\acknowledgments We would like to thank Eric Poisson and Jorge Pullin
for encouragement to study perturbations of black holes in the context
of CS gravity.  We would also like to thank Ted Jacobson, Stephon
Alexander, Daniel Grumiller, Ben Owen, and Richard O'Shaughnessy for
discussions. Most of our calculations used the computer algebra
systems MAPLE v.11 in combination with the GRTensorII
package~\cite{grtensor}.

NY would also like to thank Ben Owen for his ongoing support. CFS
acknowledges support from the Natural Sciences and Engineering
Research Council of Canada during the first stages of this work.
Presently, he is supported by the Ramon y Cajal Programme of the
Ministry of Education and Science of Spain and by a Marie Curie
International Reintegration Grant (MIRG-CT-2007-205005/PHY) within the
7th European Community Framework Programme.  NY acknowledges the
support of the Center for Gravitational Wave Physics funded by the
National Science Foundation under Cooperative Agreement PHY-01-14375,
support from NSF grant PHY-05-55-628, and the University of Guelph for
hospitality during a visit in which this work was started.

%%%%%%%%%%%%%%%%%%%%%%%%%%%%%%%%%%%%%%%%%%%%%%%%%%%%%%%%%%%%%%%%%%%%%%%%%%%%%%
\appendix

%%%%%%%%%%%%%%%%%%%%%%%%%%%%%%%%%%%%%%%%%%%%%%%%%%%%%%%%%%%%%%%%%%%%%%%%%%%%%%
\section{Explicit expressions for the perturbed C- and Einstein tensors}
\label{explicitexpressions}

Here we provide explicit expressions in Schwarzschild coordinates for
the coefficients of the expansion in spherical harmonics of the
Einstein and C-tensors. In these expressions, the Pontryagin
constraint [Eq.~(\ref{pontryagin})] has been used. For simplicity we
omit the superscripts $(\ell\,,m)$.  The components of the
harmonically decomposed Einstein tensor read
\begin{widetext}
\begin{equation}
{\cal G}^{}_{tt} = -f^2 K_{,rr} 
- f\left(\frac{f'}{2}+\frac{3f}{r}\right) K_{,r} 
+ \frac{(\ell+2)(\ell-1)f}{2r^2}K
+ \frac{f^3}{r}  h^{}_{rr,r}
+ \frac{f^2}{r^2}\left[\frac{\ell(\ell+1)}{2} + f + 2rf' \right]h^{}_{rr}\,,
\end{equation}
\begin{equation}
{\cal G}^{}_{tr} = - K_{,tr} 
+\frac{1}{2f}\left(f'-\frac{2f}{r}\right) K_{,t}
+\frac{f}{r} h^{}_{rr,t}
+\frac{\ell(\ell+1)}{2r^2}h^{}_{tr}\,,
\end{equation}
\begin{equation}
{\cal G}^{}_{rr} = -\frac{1}{f^2} K_{,tt}
+\frac{1}{2f}\left(f'+\frac{2f}{r}\right) K_{,r}
-\frac{(\ell+2)(\ell-1)}{2r^2f} K
- \frac{1}{rf} h^{}_{tt,r}
+\frac{1}{r^2f^2}\left[\frac{\ell(\ell+1)}{2}+rf'\right] h^{}_{tt}
+\frac{2}{rf}  h^{}_{tr,t} 
-\frac{1}{r^2}h^{}_{rr}\,,
\end{equation}
\begin{equation}
{\cal G}^{}_t = -\frac{1}{2} K_{,t}
+\frac{f}{2} h^{}_{tr,r}
-\frac{f}{2} h^{}_{rr,t}
+\frac{f'}{2}h^{}_{tr}\,,
\end{equation}
\begin{equation}
{\cal G}^{}_r = -\frac{1}{2} K_{,r}
+\frac{1}{2f} h^{}_{tt,r}
-\frac{1}{4f^2}\left(f'+\frac{2f}{r}\right)h^{}_{tt}
-\frac{1}{2f}  h^{}_{tr,t}
+\frac{1}{4}\left(f'+\frac{2f}{r}\right) h^{}_{rr}\,,
\end{equation}
\begin{equation}
{\cal H}^{}_t =  \frac{(\ell+2)(\ell-1)}{2r^2} h^{}_t \,,
\end{equation}
\begin{equation}
{\cal H}^{}_r =  \frac{(\ell+2)(\ell-1)}{2r^2} h^{}_r\,,
\end{equation}
\begin{eqnarray}
{\cal G} & = & -\frac{r^2}{2f} K_{,tt}
+\frac{r^2 f}{2} K_{,rr}
+\frac{r^2}{2}\left(f' +\frac{2f}{r}\right) K_{,r}
-\frac{r^2}{2} h^{}_{tt,rr}
+\frac{r^2}{4f}\left(f' - \frac{2f}{r}\right) h^{}_{tt,r}
\nn \\
& + & \frac{r^2}{2f}\left[ \frac{\ell(\ell+1)}{2r^2} + f'' + \frac{f'}{r}
                     -\frac{f'^2}{2f}\right]h^{}_{tt}
+r^2  h^{}_{tr,tr}
+\frac{r^2}{2f}\left( f' + \frac{2f}{r}\right) h^{}_{tr,t}
-\frac{r^2}{2} h^{}_{rr,tt} \nn \\
& - & \frac{r^2 f}{4}\left(f' + \frac{2f}{r}\right) h^{}_{rr,r}
-\frac{r^2f}{2}\left[ \frac{\ell(\ell+1)}{2r^2} + f'' +\frac{3f'}{r}
                     +\frac{f'^2}{2f}\right] h^{}_{rr}\,,
\end{eqnarray}
\begin{equation}
{\cal H} = \frac{1}{2f}\left( h^{}_{tt} - f^2 h^{}_{rr} \right)\,,
\end{equation}
\begin{equation}
{\cal I} = -\frac{1}{f}\left( h^{}_{t,t} - f^2 
h^{}_{r,r} -ff'h^{}_r\right)\,.
\end{equation}
\end{widetext}
while the components of the harmonically decomposed C-tensor read
\begin{widetext}
\begin{equation}
{\cal C}^{}_{tt} = -\frac{(l+2)!}{(l-2)!} \frac{f}{2 r^4} \bar{\theta}_{,r} h^{}_t\,,
\end{equation}
\begin{equation}
{\cal C}^{}_{tr} = - \frac{(\ell+2)!}{(\ell-2)!} \frac{1}{4r^4f}
\left( \bar{\theta}^{}_{,t} h^{}_t +  f^2 \bar{\theta}^{}_{,r} h^{}_r\right)\,,
\label{Ctr}
\end{equation}
\begin{equation}
{\cal C}^{}_{rr} = -\frac{(\ell+2)!}{(\ell-2)!}
\frac{\bar{\theta}_{,t}}{2r^4f}h^{}_r\,, 
\label{Crr}
\end{equation}
\begin{equation}
{\cal C}^{}_t = - \frac{(\ell+2)(\ell-1)}{4r^2} \left\{
\bar{\theta}_{,t} \left(  h^{}_{t,r} - \frac{2 h^{}_t}{r} \right) 
+ f^2 \bar{\theta}_{,r} \left[  h_{r,r} - \frac{2 h_r}{r f} \left(1 - \frac{4
      M}{r} \right) - \frac{2  h_{t,t}}{f^2} \right] 
+ f^2 \bar{\theta}_{,rr} h_r - \bar{\theta}_{,tr} h_t \right\}\,,
\end{equation}
\begin{equation}
{\cal C}^{}_r = - \frac{\left(\ell+2\right) \left(\ell - 1\right)}{4
  r^3 f^2} \left[
\bar{\theta}_{,t} \left(-r  h_{t,t} + 2 r f^2
 h_{r,r} + r f f' h_r - 2 f^2 h_r \right)
- \bar{\theta}_{,r} \left(-r f^2  h_{t,r} + r f f' h_t \right)
- \bar{\theta}_{,tt} r h_t+ \bar{\theta}_{,tr} f^2 r h_r \right]\,,
\end{equation}
\begin{eqnarray}
{\cal D}^{}_t &=& \bar{\theta}_{,t} \left( \frac{1}{4} K_{,tr}
- \frac{1}{4 f}\left(f'-\frac{2f}{r}\right) K_{,t}
+ \frac{f}{4}  h^{}_{tr,rr}
+ \frac{f'}{4}  h^{}_{tr,r} - \frac{f}{2 r}
 h^{}_{rr,t} 
\right. 
\nn \\
&-& \left. 
\frac{1}{4} \left\{ - f'' + \frac{f'^2}{f} +
  \frac{2}{r^2} \left[\frac{\ell}{2} \left(\ell+1\right) + f - 1 - r
    f' \right] \right\}h^{}_{tr} - \frac{f}{4} 
h^{}_{rr,tr} \right)
\nn \\
&+& \bar{\theta}_{,r} \left\{ \frac{K f}{4 r^2} \left(\ell +2 \right)
  \left( \ell -1 
  \right) + \frac{h_{tt}}{8 r^2 f} \left[ 2 f \ell \left( \ell + 1 \right)
    -1 - 2 f + f^2 \right] - \frac{h_{rr} f}{8 r^2} \left(1 - 6 f + 3
    f^2 \right) 
\right. 
\nn \\
&-& \left. 
 \frac{1 - 3 f}{8 r} \left( f^2  h_{rr,r} -
 h_{tt,r} \right) - \frac{f}{4} \left(  h_{tt,rr}
    - \frac{2}{r}  h_{tr,t} -  h_{tr,tr} + f
 K_{,rr} \right) - \frac{f}{4 r}  K_{,r} \left(1 +
    f\right) \right\}
\nn \\
&+& \bar{\theta}_{,rr} \left\{ \frac{1}{8 r} \left[ \left(3 f - 1\right) h_{rr}
    f^2 + \left(1 + f\right) h_{tt} \right] - \frac{f}{4} \left(
 h_{tt,r} -  h_{tr,t} + f  K_{,r} \right)
\right\} 
\nn \\
&+& \bar{\theta}_{,tr} \left[ \frac{1}{4}  K_{,t} - \frac{f}{4}
 h_{rr,t} + \frac{f}{4}  h_{tr,r} - \frac{h_{tr}}{4r}
  \left(3 f - 1 \right) \right]
- \frac{f f'''}{4} \tilde{\theta} + \frac{3 M}{r^3} f 
\tilde{\theta}_{,r} \,,
\end{eqnarray}
\begin{eqnarray}
{\cal D}^{}_r &=& 
\frac{\bar{\theta}_{,t}}{4 f} \left\{ -  h_{rr,tt} +
  \frac{1}{r^2} \left[ \ell \left(\ell+1 \right) - 2
  \right] \left(K - f h_{rr} \right) +  h_{tr,t} 
  \left( - \frac{2}{r} + \frac{f'}{f} \right) 
+  h_{tr,tr} + \frac{1}{f}  K_{,tt} \right\}
\nn \\
&+& \bar{\theta}_{,r} \left[ -\frac{1}{8r} \left(f' r - 2  f\right) 
  h_{rr,t} + \frac{1}{8 f^2 r} \left(f'r  + 2 f\right) 
  h_{tt,t} + \frac{1}{4f} \left( h_{tr,tt} - 
  h_{tt,tr} - f K_{,tr} \right) + \frac{\left(\ell+2\right)
  \left(\ell -1\right)}{4 r^2} h_{tr} \right]
\nn \\
&-& \frac{\bar{\theta}_{,tt}}{4 r f^2} \left( - r f h_{tr,r}
  + 2 h_{tr} f + r f  h_{rr,t} - r f' h_{tr} - r  K_{,t}
  \right) 
\nn \\
&+& \bar{\theta}_{,tr} \left[ \frac{1}{4 f} \left(  h_{tr,t} - f
 K_{,r} -  h_{tt,r}\right) + \frac{h_{tt}}{8 r f^2} \left(r
  f' + 2 f \right) - \frac{h_{rr}}{8 r} \left( r f' - 2 f \right) \right]
+ \frac{3 M}{f r^3} \tilde{\theta}_{,t}\,,
\end{eqnarray}
\begin{eqnarray}
{\cal C} = \frac{(\ell+2)!}{(\ell-2)!} \frac{1}{4r^2} 
\left(\bar{\theta}_{,t} h^{}_r  - \bar{\theta}_{,r} h_t \right),
\end{eqnarray}
\begin{eqnarray}
{\cal D} & = & 
\bar{\theta}_{,t} \left\{
- f h^{}_{r,rr} - \left(\frac{3}{2} f' - \frac{f}{r}\right)
 h^{}_{r,r} + \frac{1}{2} \left[\frac{\ell(\ell+1)}{r^2} -
    f'' + \frac{2}{r}\left(f'-\frac{f}{r}\right) \right] h^{}_r 
+ \frac{1}{f} h^{}_{t,tr} -
\frac{1}{2f^2}\left(f'+\frac{2f}{r}\right) h^{}_{t,t}  
\right\}
\nn \\
&+& \bar{\theta}_{,r} \left\{ -\frac{1}{f} \left(h_{t,tt} - f^2
 h_{t,rr} \right) - \frac{h_t}{2 r^2} \left[4 f' r + f''
    r^2 - 6 f + \ell \left( \ell + 1\right) \right] + \frac{1}{r}
  \left(f' r - 2 f \right)  h_{t,r} \right\} 
\nn \\
&+& \left(\bar{\theta}_{,rr}+ \frac{\bar{\theta}_{,tt}}{f^2} \right) \left[-
  \frac{h_t}{2 r} \left(2 f + f' r\right) + f  h_{t,r} \right]
- \bar{\theta}_{,tr} \left( f  h_{r,r} + \frac{ h_{t,t}}{f} \right) \,,
\end{eqnarray}
\begin{eqnarray}
{\cal E} &=& \bar{\theta}_{,t} \left[
\frac{1}{2} K_{,r} + \frac{1}{2 f}h^{}_{tr,t} 
- \frac{f}{2} h^{}_{rr,r} - \frac{f'}{2} h^{}_{rr} \right] 
+ \frac{\bar{\theta}_{,r}}{2f} \left( f^2  h_{tr,r} - f K_{,t} -
 h_{tt,t} + f f' h_{tr} \right)
\nn \\
&+& \frac{f}{2} h_{tr} \left(\bar{\theta}_{,rr}  + \frac{\bar{\theta}_{,tt}}{f^2} \right) 
- \frac{\bar{\theta}_{,tr}}{2 f} \left(h_{tt} + f^2 h_{rr} \right)\,.
\end{eqnarray}
\end{widetext}

\section{Equations of the extended CS theory in the Regge-Wheeler gauge}
\label{rwexpressions}

The equation for the background component of the scalar field, 
Eq.~(\ref{eqbartheta}), is
\be
\left(-\partial^2_{t^2} + \partial^2_{r^{2}_\star} -\frac{2Mf}{r^3}\right)
(r\,\bar\theta) = 0\,,
\ee

The divergence of the field equations, which in the non-extended
theory leads to the Pontryagin constraint, now leads to the equations
of motion for the scalar field. To leading order in $\epsilon$, this
equation reduces to Eq.~(\ref{eqdeltatheta}), which when harmonically
decomposed becomes 
\begin{widetext}
\ba
&& \left[f \tilde{\theta}^{\ell m}_{,rr} - \frac{1}{f}
   \tilde{\theta}^{\ell m}_{,tt} + \frac{2}{r} \left(1 -
  \frac{M}{r} \right) \tilde{\theta}^{\ell m}_{,r} 
-  \frac{\tilde{\theta}^{\ell m}}{r^2} \ell \left(  
  \ell + 1 \right) \right]
- \left[ f^2 h_{rr}^{\ell m} \bar{\theta}_{,rr} + \frac{1}{f^2}
    h_{tt}^{\ell m} \bar{\theta}_{,tt} \right]
\nonumber \\
&+& \left(-\frac{1}{2 f^2} h_{tt,t}^{\ell m} -
  \frac{1}{2} h_{rr,t}^{\ell m} +  h_{tr,r}^{\ell m}
  + \frac{2}{r} h_{tr}^{\ell m} \right) \bar{\theta}_{,t}
  + \left[ -\frac{f^2}{2} h_{rr,r}^{\ell m} - \frac{1}{2}
 h_{tt,r}^{\ell m} 
\right. 
\nonumber \\
&+& \left.  h_{tr,t}^{\ell m} +
  \frac{M}{r^2 f} h_{tt}^{\ell m} - \frac{f}{2 r} \left(3 + f\right)
  h_{rr}^{\ell m} \right]  \bar{\theta}_{,r} + f  K^{\ell m}_{,r}
 \bar{\theta}_{,r} 
- \frac{ K^{\ell m}_{,t}}{f} \bar{\theta}_{,t} 
= \frac{6 M}{r^6} \frac{\left(\ell + 2 \right)!}{\left(\ell -
    2 \right)!} \Psi_{\CPM}^{\ell m}\,.
\label{eqdeltathetarw}
\ea
\end{widetext}
%

%%%%%%%%%%%%%%%%%%%%%%%%%%%%%%%%%%%%%%%%%%%%%%%%%%%%%%%%%%%%%%%%%%%%%%%%%%%%%%
\bibliography{phyjabb,master}

\begin{thebibliography}{58}
\expandafter\ifx\csname natexlab\endcsname\relax\def\natexlab#1{#1}\fi
\expandafter\ifx\csname bibnamefont\endcsname\relax
  \def\bibnamefont#1{#1}\fi
\expandafter\ifx\csname bibfnamefont\endcsname\relax
  \def\bibfnamefont#1{#1}\fi
\expandafter\ifx\csname citenamefont\endcsname\relax
  \def\citenamefont#1{#1}\fi
\expandafter\ifx\csname url\endcsname\relax
  \def\url#1{\texttt{#1}}\fi
\expandafter\ifx\csname urlprefix\endcsname\relax\def\urlprefix{URL }\fi
\providecommand{\bibinfo}[2]{#2}
\providecommand{\eprint}[2][]{\url{#2}}

\bibitem[{\citenamefont{Jackiw and Pi}(2003)}]{jackiw:2003:cmo}
\bibinfo{author}{\bibfnamefont{R.}~\bibnamefont{Jackiw}} \bibnamefont{and}
  \bibinfo{author}{\bibfnamefont{S.~Y.} \bibnamefont{Pi}},
  \bibinfo{journal}{Phys. Rev.} \textbf{\bibinfo{volume}{D68}},
  \bibinfo{pages}{104012} (\bibinfo{year}{2003}), \eprint{gr-qc/0308071}.

\bibitem[{\citenamefont{Alexander et~al.}(2006)\citenamefont{Alexander, Peskin,
  and Sheik-Jabbari}}]{alexander:2004:lfg}
\bibinfo{author}{\bibfnamefont{S.~H.~S.} \bibnamefont{Alexander}},
  \bibinfo{author}{\bibfnamefont{M.~E.} \bibnamefont{Peskin}},
  \bibnamefont{and} \bibinfo{author}{\bibfnamefont{M.~M.}
  \bibnamefont{Sheik-Jabbari}}, \bibinfo{journal}{Phys. Rev. Lett.}
  \textbf{\bibinfo{volume}{96}}, \bibinfo{pages}{081301}
  (\bibinfo{year}{2006}), \eprint{hep-th/0403069}.

\bibitem[{\citenamefont{Lue et~al.}(1999)\citenamefont{Lue, Wang, and
  Kamionkowski}}]{Lue:1998mq}
\bibinfo{author}{\bibfnamefont{A.}~\bibnamefont{Lue}},
  \bibinfo{author}{\bibfnamefont{L.-M.} \bibnamefont{Wang}}, \bibnamefont{and}
  \bibinfo{author}{\bibfnamefont{M.}~\bibnamefont{Kamionkowski}},
  \bibinfo{journal}{Phys. Rev. Lett.} \textbf{\bibinfo{volume}{83}},
  \bibinfo{pages}{1506} (\bibinfo{year}{1999}), \eprint{astro-ph/9812088}.

\bibitem[{\citenamefont{Alexander
  et~al.}(2007{\natexlab{a}})\citenamefont{Alexander, Finn, and
  Yunes}}]{Alexander:2007:bgw}
\bibinfo{author}{\bibfnamefont{S.}~\bibnamefont{Alexander}},
  \bibinfo{author}{\bibfnamefont{L.~S.} \bibnamefont{Finn}}, \bibnamefont{and}
  \bibinfo{author}{\bibfnamefont{N.}~\bibnamefont{Yunes}}, \bibinfo{journal}{in
  progress}  (\bibinfo{year}{2007}{\natexlab{a}}).

\bibitem[{\citenamefont{Alexander}(2007)}]{stephon}
\bibinfo{author}{\bibfnamefont{S.}~\bibnamefont{Alexander}},
  \bibinfo{journal}{in progress}  (\bibinfo{year}{2007}).

\bibitem[{\citenamefont{Green et~al.}(1987)\citenamefont{Green, Schwarz, and
  Witten}}]{Green:1987mn}
\bibinfo{author}{\bibfnamefont{M.~B.} \bibnamefont{Green}},
  \bibinfo{author}{\bibfnamefont{J.~H.} \bibnamefont{Schwarz}},
  \bibnamefont{and} \bibinfo{author}{\bibfnamefont{E.}~\bibnamefont{Witten}},
  \emph{\bibinfo{title}{Superstring Theory. Vol. 2: Loop Amplitides, Anomalies
  and Phenomenology}} (\bibinfo{publisher}{Cambridge University Press},
  \bibinfo{address}{Cambridge, UK}, \bibinfo{year}{1987}).

\bibitem[{\citenamefont{Polchinski}(1998)}]{Polchinski:1998rr}
\bibinfo{author}{\bibfnamefont{J.}~\bibnamefont{Polchinski}},
  \emph{\bibinfo{title}{String theory. Vol. 2: Superstring theory and beyond}}
  (\bibinfo{publisher}{Cambridge University Press},
  \bibinfo{address}{Cambridge, UK}, \bibinfo{year}{1998}).

\bibitem[{\citenamefont{Li et~al.}(2006)\citenamefont{Li, Xia, Li, and
  Zhang}}]{Li:2006ss}
\bibinfo{author}{\bibfnamefont{M.}~\bibnamefont{Li}},
  \bibinfo{author}{\bibfnamefont{J.-Q.} \bibnamefont{Xia}},
  \bibinfo{author}{\bibfnamefont{H.}~\bibnamefont{Li}}, \bibnamefont{and}
  \bibinfo{author}{\bibfnamefont{X.}~\bibnamefont{Zhang}}
  (\bibinfo{year}{2006}), \eprint{hep-ph/0611192}.

\bibitem[{\citenamefont{Alexander}(2006)}]{Alexander:2006mt}
\bibinfo{author}{\bibfnamefont{S.~H.~S.} \bibnamefont{Alexander}}
  (\bibinfo{year}{2006}), \eprint{hep-th/0601034}.

\bibitem[{\citenamefont{Alexander and Gates}(2006)}]{Alexander:2004xd}
\bibinfo{author}{\bibfnamefont{S.~H.~S.} \bibnamefont{Alexander}}
  \bibnamefont{and} \bibinfo{author}{\bibfnamefont{J.}~\bibnamefont{Gates},
  \bibfnamefont{S.~James}}, \bibinfo{journal}{JCAP}
  \textbf{\bibinfo{volume}{0606}}, \bibinfo{pages}{018} (\bibinfo{year}{2006}),
  \eprint{hep-th/0409014}.

\bibitem[{\citenamefont{Alexander and
  Martin}(2005{\natexlab{a}})}]{Alexander:2004wk}
\bibinfo{author}{\bibfnamefont{S.}~\bibnamefont{Alexander}} \bibnamefont{and}
  \bibinfo{author}{\bibfnamefont{J.}~\bibnamefont{Martin}},
  \bibinfo{journal}{Phys. Rev.} \textbf{\bibinfo{volume}{D71}},
  \bibinfo{pages}{063526} (\bibinfo{year}{2005}{\natexlab{a}}),
  \eprint{hep-th/0410230}.

\bibitem[{\citenamefont{Alexander and
  Yunes}(2007{\natexlab{a}})}]{Alexander:2007zg}
\bibinfo{author}{\bibfnamefont{S.}~\bibnamefont{Alexander}} \bibnamefont{and}
  \bibinfo{author}{\bibfnamefont{N.}~\bibnamefont{Yunes}}
  (\bibinfo{year}{2007}{\natexlab{a}}), \eprint{hep-th/0703265}.

\bibitem[{\citenamefont{Alexander and
  Yunes}(2007{\natexlab{b}})}]{Alexander:2007vt}
\bibinfo{author}{\bibfnamefont{S.}~\bibnamefont{Alexander}} \bibnamefont{and}
  \bibinfo{author}{\bibfnamefont{N.}~\bibnamefont{Yunes}},
  \bibinfo{journal}{Phys. Rev.} \textbf{\bibinfo{volume}{D75}},
  \bibinfo{pages}{124022} (\bibinfo{year}{2007}{\natexlab{b}}),
  \eprint{arXiv:0704.0299 [hep-th]}.

\bibitem[{\citenamefont{Smith et~al.}(2007)\citenamefont{Smith, Erickcek,
  Caldwell, and Kamionkowski}}]{Smith:2007jm}
\bibinfo{author}{\bibfnamefont{T.~L.} \bibnamefont{Smith}},
  \bibinfo{author}{\bibfnamefont{A.~L.} \bibnamefont{Erickcek}},
  \bibinfo{author}{\bibfnamefont{R.~R.} \bibnamefont{Caldwell}},
  \bibnamefont{and}
  \bibinfo{author}{\bibfnamefont{M.}~\bibnamefont{Kamionkowski}}
  (\bibinfo{year}{2007}), \eprint{arXiv:0708.0001 [astro-ph]}.

\bibitem[{\citenamefont{Kostelecky}(2004)}]{Kostelecky:2003fs}
\bibinfo{author}{\bibfnamefont{V.~A.} \bibnamefont{Kostelecky}},
  \bibinfo{journal}{Phys. Rev.} \textbf{\bibinfo{volume}{D69}},
  \bibinfo{pages}{105009} (\bibinfo{year}{2004}), \eprint{hep-th/0312310}.

\bibitem[{\citenamefont{Mariz et~al.}(2004)\citenamefont{Mariz, Nascimento,
  Passos, and Ribeiro}}]{Mariz:2004cv}
\bibinfo{author}{\bibfnamefont{T.}~\bibnamefont{Mariz}},
  \bibinfo{author}{\bibfnamefont{J.~R.} \bibnamefont{Nascimento}},
  \bibinfo{author}{\bibfnamefont{E.}~\bibnamefont{Passos}}, \bibnamefont{and}
  \bibinfo{author}{\bibfnamefont{R.~F.} \bibnamefont{Ribeiro}},
  \bibinfo{journal}{Phys. Rev.} \textbf{\bibinfo{volume}{D70}},
  \bibinfo{pages}{024014} (\bibinfo{year}{2004}), \eprint{hep-th/0403205}.

\bibitem[{\citenamefont{Bluhm and Kostelecky}(2005)}]{Bluhm:2004ep}
\bibinfo{author}{\bibfnamefont{R.}~\bibnamefont{Bluhm}} \bibnamefont{and}
  \bibinfo{author}{\bibfnamefont{V.~A.} \bibnamefont{Kostelecky}},
  \bibinfo{journal}{Phys. Rev.} \textbf{\bibinfo{volume}{D71}},
  \bibinfo{pages}{065008} (\bibinfo{year}{2005}), \eprint{hep-th/0412320}.

\bibitem[{\citenamefont{Eling et~al.}(2004)\citenamefont{Eling, Jacobson, and
  Mattingly}}]{Eling:2004dk}
\bibinfo{author}{\bibfnamefont{C.}~\bibnamefont{Eling}},
  \bibinfo{author}{\bibfnamefont{T.}~\bibnamefont{Jacobson}}, \bibnamefont{and}
  \bibinfo{author}{\bibfnamefont{D.}~\bibnamefont{Mattingly}}
  (\bibinfo{year}{2004}), \eprint{gr-qc/0410001}.

\bibitem[{\citenamefont{Lyth et~al.}(2005)\citenamefont{Lyth, Quimbay, and
  Rodriguez}}]{Lyth:2005jf}
\bibinfo{author}{\bibfnamefont{D.~H.} \bibnamefont{Lyth}},
  \bibinfo{author}{\bibfnamefont{C.}~\bibnamefont{Quimbay}}, \bibnamefont{and}
  \bibinfo{author}{\bibfnamefont{Y.}~\bibnamefont{Rodriguez}},
  \bibinfo{journal}{JHEP} \textbf{\bibinfo{volume}{03}}, \bibinfo{pages}{016}
  (\bibinfo{year}{2005}), \eprint{hep-th/0501153}.

\bibitem[{\citenamefont{Mattingly}(2005)}]{Mattingly:2005re}
\bibinfo{author}{\bibfnamefont{D.}~\bibnamefont{Mattingly}},
  \bibinfo{journal}{Living Rev. Rel.} \textbf{\bibinfo{volume}{8}},
  \bibinfo{pages}{5} (\bibinfo{year}{2005}), \eprint{gr-qc/0502097}.

\bibitem[{\citenamefont{Lehnert}(2006)}]{Lehnert:2006rp}
\bibinfo{author}{\bibfnamefont{R.}~\bibnamefont{Lehnert}}
  (\bibinfo{year}{2006}), \eprint{gr-qc/0602073}.

\bibitem[{\citenamefont{Hariton and Lehnert}(2007)}]{Hariton:2006zj}
\bibinfo{author}{\bibfnamefont{A.~J.} \bibnamefont{Hariton}} \bibnamefont{and}
  \bibinfo{author}{\bibfnamefont{R.}~\bibnamefont{Lehnert}},
  \bibinfo{journal}{Phys. Lett.} \textbf{\bibinfo{volume}{A367}},
  \bibinfo{pages}{11} (\bibinfo{year}{2007}), \eprint{hep-th/0612167}.

\bibitem[{\citenamefont{Alexander
  et~al.}(2007{\natexlab{b}})\citenamefont{Alexander, Peskin, and
  Sheikh-Jabbari}}]{Alexander:2007qe}
\bibinfo{author}{\bibfnamefont{S.~H.} \bibnamefont{Alexander}},
  \bibinfo{author}{\bibfnamefont{M.~E.} \bibnamefont{Peskin}},
  \bibnamefont{and} \bibinfo{author}{\bibfnamefont{M.~M.}
  \bibnamefont{Sheikh-Jabbari}} (\bibinfo{year}{2007}{\natexlab{b}}),
  \eprint{hep-ph/0701139}.

\bibitem[{\citenamefont{Guarrera and Hariton}(2007)}]{Guarrera:2007tu}
\bibinfo{author}{\bibfnamefont{D.}~\bibnamefont{Guarrera}} \bibnamefont{and}
  \bibinfo{author}{\bibfnamefont{A.~J.} \bibnamefont{Hariton}}
  (\bibinfo{year}{2007}), \eprint{gr-qc/0702029}.

\bibitem[{\citenamefont{Konno et~al.}(2007)\citenamefont{Konno, Matsuyama, and
  Tanda}}]{Konno:2007ze}
\bibinfo{author}{\bibfnamefont{K.}~\bibnamefont{Konno}},
  \bibinfo{author}{\bibfnamefont{T.}~\bibnamefont{Matsuyama}},
  \bibnamefont{and} \bibinfo{author}{\bibfnamefont{S.}~\bibnamefont{Tanda}},
  \bibinfo{journal}{Phys. Rev.} \textbf{\bibinfo{volume}{D76}},
  \bibinfo{pages}{024009} (\bibinfo{year}{2007}), \eprint{arXiv:0706.3080
  [gr-qc]}.

\bibitem[{\citenamefont{Fischler and Paban}(2007)}]{Fischler:2007tj}
\bibinfo{author}{\bibfnamefont{W.}~\bibnamefont{Fischler}} \bibnamefont{and}
  \bibinfo{author}{\bibfnamefont{S.}~\bibnamefont{Paban}}
  (\bibinfo{year}{2007}), \eprint{arXiv:0708.3828 [hep-th]}.

\bibitem[{\citenamefont{Tekin}(2007)}]{Tekin:2007rn}
\bibinfo{author}{\bibfnamefont{B.}~\bibnamefont{Tekin}} (\bibinfo{year}{2007}),
  \eprint{arXiv:0710.2528 [gr-qc]}.

\bibitem[{\citenamefont{Grumiller and Jackiw}(2007)}]{Grumiller:2007gm}
\bibinfo{author}{\bibfnamefont{D.}~\bibnamefont{Grumiller}} \bibnamefont{and}
  \bibinfo{author}{\bibfnamefont{R.}~\bibnamefont{Jackiw}}
  (\bibinfo{year}{2007}), \eprint{arXiv:0711.0181 [math-ph]}.

\bibitem[{\citenamefont{Grumiller and Yunes}(2007)}]{Grumiller:2007rv}
\bibinfo{author}{\bibfnamefont{D.}~\bibnamefont{Grumiller}} \bibnamefont{and}
  \bibinfo{author}{\bibfnamefont{N.}~\bibnamefont{Yunes}}
  (\bibinfo{year}{2007}), \eprint{arXiv:0711.1868 [gr-qc]}.

\bibitem[{\citenamefont{Gerlach and Sengupta}(1979)}]{Gerlach:1979rw}
\bibinfo{author}{\bibfnamefont{U.~H.} \bibnamefont{Gerlach}} \bibnamefont{and}
  \bibinfo{author}{\bibfnamefont{U.~K.} \bibnamefont{Sengupta}},
  \bibinfo{journal}{Phys. Rev.} \textbf{\bibinfo{volume}{D19}},
  \bibinfo{pages}{2268} (\bibinfo{year}{1979}).

\bibitem[{\citenamefont{Gerlach and Sengupta}(1980)}]{Gerlach:1980tx}
\bibinfo{author}{\bibfnamefont{U.~H.} \bibnamefont{Gerlach}} \bibnamefont{and}
  \bibinfo{author}{\bibfnamefont{U.~K.} \bibnamefont{Sengupta}},
  \bibinfo{journal}{Phys. Rev.} \textbf{\bibinfo{volume}{D22}},
  \bibinfo{pages}{1300} (\bibinfo{year}{1980}).

\bibitem[{\citenamefont{Sopuerta et~al.}(2006)\citenamefont{Sopuerta, Yunes,
  and Laguna}}]{Sopuerta:2006wj}
\bibinfo{author}{\bibfnamefont{C.~F.} \bibnamefont{Sopuerta}},
  \bibinfo{author}{\bibfnamefont{N.}~\bibnamefont{Yunes}}, \bibnamefont{and}
  \bibinfo{author}{\bibfnamefont{P.}~\bibnamefont{Laguna}},
  \bibinfo{journal}{Phys. Rev.} \textbf{\bibinfo{volume}{D74}},
  \bibinfo{pages}{124010} (\bibinfo{year}{2006}), \eprint{astro-ph/0608600}.

\bibitem[{\citenamefont{{Cunningham} et~al.}(1978)\citenamefont{{Cunningham},
  {Price}, and {Moncrief}}}]{Cunningham:1978cp}
\bibinfo{author}{\bibfnamefont{C.~T.} \bibnamefont{{Cunningham}}},
  \bibinfo{author}{\bibfnamefont{R.~H.} \bibnamefont{{Price}}},
  \bibnamefont{and}
  \bibinfo{author}{\bibfnamefont{V.}~\bibnamefont{{Moncrief}}},
  \bibinfo{journal}{\apj} \textbf{\bibinfo{volume}{224}}, \bibinfo{pages}{643}
  (\bibinfo{year}{1978}).

\bibitem[{\citenamefont{Zerilli}(1970)}]{Zerilli:1970fj}
\bibinfo{author}{\bibfnamefont{F.~J.} \bibnamefont{Zerilli}},
  \bibinfo{journal}{\prl} \textbf{\bibinfo{volume}{24}}, \bibinfo{pages}{737}
  (\bibinfo{year}{1970}).

\bibitem[{\citenamefont{Moncrief}(1974)}]{Moncrief:1974vm}
\bibinfo{author}{\bibfnamefont{V.}~\bibnamefont{Moncrief}},
  \bibinfo{journal}{Ann. Phys. (N.Y.)} \textbf{\bibinfo{volume}{88}},
  \bibinfo{pages}{323} (\bibinfo{year}{1974}).

\bibitem[{\citenamefont{Goldberg et~al.}(1967)\citenamefont{Goldberg,
  Macfarlane, Newman, Rohrlich, and Sudarshan}}]{Goldberg:1967sp}
\bibinfo{author}{\bibfnamefont{J.~N.} \bibnamefont{Goldberg}},
  \bibinfo{author}{\bibfnamefont{A.~J.} \bibnamefont{Macfarlane}},
  \bibinfo{author}{\bibfnamefont{E.~T.} \bibnamefont{Newman}},
  \bibinfo{author}{\bibfnamefont{F.}~\bibnamefont{Rohrlich}}, \bibnamefont{and}
  \bibinfo{author}{\bibfnamefont{E.~C.~G.} \bibnamefont{Sudarshan}},
  \bibinfo{journal}{J. Math. Phys.} \textbf{\bibinfo{volume}{8}},
  \bibinfo{pages}{2155} (\bibinfo{year}{1967}).

\bibitem[{\citenamefont{{Isaacson}}(1968{\natexlab{a}})}]{Isaacson:1968ra}
\bibinfo{author}{\bibfnamefont{R.~A.} \bibnamefont{{Isaacson}}},
  \bibinfo{journal}{Phys. Rev.} \textbf{\bibinfo{volume}{166}},
  \bibinfo{pages}{1263} (\bibinfo{year}{1968}{\natexlab{a}}).

\bibitem[{\citenamefont{{Isaacson}}(1968{\natexlab{b}})}]{Isaacson:1968gw}
\bibinfo{author}{\bibfnamefont{R.~A.} \bibnamefont{{Isaacson}}},
  \bibinfo{journal}{Phys. Rev.} \textbf{\bibinfo{volume}{166}},
  \bibinfo{pages}{1272} (\bibinfo{year}{1968}{\natexlab{b}}).

\bibitem[{\citenamefont{Misner et~al.}(1973)\citenamefont{Misner, Thorne, and
  Wheeler}}]{Misner:1973cw}
\bibinfo{author}{\bibfnamefont{C.~W.} \bibnamefont{Misner}},
  \bibinfo{author}{\bibfnamefont{K.}~\bibnamefont{Thorne}}, \bibnamefont{and}
  \bibinfo{author}{\bibfnamefont{J.~A.} \bibnamefont{Wheeler}},
  \emph{\bibinfo{title}{Gravitation}} (\bibinfo{publisher}{W. H. Freeman \&
  Co.}, \bibinfo{address}{San Francisco}, \bibinfo{year}{1973}).

\bibitem[{\citenamefont{Alexander and
  Martin}(2005{\natexlab{b}})}]{alexander:2005:bgw}
\bibinfo{author}{\bibfnamefont{S.}~\bibnamefont{Alexander}} \bibnamefont{and}
  \bibinfo{author}{\bibfnamefont{J.}~\bibnamefont{Martin}},
  \bibinfo{journal}{Phys. Rev.} \textbf{\bibinfo{volume}{D71}},
  \bibinfo{pages}{063526} (\bibinfo{year}{2005}{\natexlab{b}}),
  \eprint{hep-th/0410230}.

\bibitem[{grt()}]{grtensor}
\emph{\bibinfo{title}{Grtensorii}}, \bibinfo{note}{this is a package which runs
  within Maple but distinct from packages distributed with Maple. It is
  distributed freely on the World-Wide-Web from the address: {\tt
  http://grtensor.org}}.

\bibitem[{\citenamefont{Sopuerta}(1997)}]{Sopuerta:1997nh}
\bibinfo{author}{\bibfnamefont{C.~F.} \bibnamefont{Sopuerta}},
  \bibinfo{journal}{Phys. Rev.} \textbf{\bibinfo{volume}{D55}},
  \bibinfo{pages}{5936} (\bibinfo{year}{1997}).

\bibitem[{\citenamefont{Sopuerta et~al.}(1999)\citenamefont{Sopuerta, Maartens,
  Ellis, and Lesame}}]{Sopuerta:1998rt}
\bibinfo{author}{\bibfnamefont{C.~F.} \bibnamefont{Sopuerta}},
  \bibinfo{author}{\bibfnamefont{R.}~\bibnamefont{Maartens}},
  \bibinfo{author}{\bibfnamefont{G.~F.~R.} \bibnamefont{Ellis}},
  \bibnamefont{and} \bibinfo{author}{\bibfnamefont{W.~M.}
  \bibnamefont{Lesame}}, \bibinfo{journal}{Phys. Rev.}
  \textbf{\bibinfo{volume}{D60}}, \bibinfo{pages}{024006}
  (\bibinfo{year}{1999}), \eprint{gr-qc/9809085}.

\bibitem[{\citenamefont{Carrol}(2003)}]{Carrol}
\bibinfo{author}{\bibfnamefont{S.~M.} \bibnamefont{Carrol}},
  \emph{\bibinfo{title}{An introduction to general relativity, Spacetime and
  Geometry}} (\bibinfo{publisher}{Pearson - Benjamin Cummings},
  \bibinfo{address}{San Francisco}, \bibinfo{year}{2003}).

\bibitem[{\citenamefont{Svrcek and Witten}(2006)}]{Svrcek:2006yi}
\bibinfo{author}{\bibfnamefont{P.}~\bibnamefont{Svrcek}} \bibnamefont{and}
  \bibinfo{author}{\bibfnamefont{E.}~\bibnamefont{Witten}},
  \bibinfo{journal}{JHEP} \textbf{\bibinfo{volume}{06}}, \bibinfo{pages}{051}
  (\bibinfo{year}{2006}), \eprint{hep-th/0605206}.

\bibitem[{\citenamefont{Brandenberger and Vafa}(1989)}]{Brandenberger:1988:sit}
\bibinfo{author}{\bibfnamefont{R.~H.} \bibnamefont{Brandenberger}}
  \bibnamefont{and} \bibinfo{author}{\bibfnamefont{C.}~\bibnamefont{Vafa}},
  \bibinfo{journal}{Nucl. Phys.} \textbf{\bibinfo{volume}{B316}},
  \bibinfo{pages}{391} (\bibinfo{year}{1989}).

\bibitem[{\citenamefont{Tseytlin and Vafa}(1992)}]{Tseytlin:1991:eos}
\bibinfo{author}{\bibfnamefont{A.~A.} \bibnamefont{Tseytlin}} \bibnamefont{and}
  \bibinfo{author}{\bibfnamefont{C.}~\bibnamefont{Vafa}},
  \bibinfo{journal}{Nucl. Phys.} \textbf{\bibinfo{volume}{B372}},
  \bibinfo{pages}{443} (\bibinfo{year}{1992}), \eprint{hep-th/9109048}.

\bibitem[{\citenamefont{Nayeri et~al.}(2006)\citenamefont{Nayeri,
  Brandenberger, and Vafa}}]{Nayeri:2005:pas}
\bibinfo{author}{\bibfnamefont{A.}~\bibnamefont{Nayeri}},
  \bibinfo{author}{\bibfnamefont{R.~H.} \bibnamefont{Brandenberger}},
  \bibnamefont{and} \bibinfo{author}{\bibfnamefont{C.}~\bibnamefont{Vafa}},
  \bibinfo{journal}{Phys. Rev. Lett.} \textbf{\bibinfo{volume}{97}},
  \bibinfo{pages}{021302} (\bibinfo{year}{2006}), \eprint{hep-th/0511140}.

\bibitem[{\citenamefont{Sun and Zhang}(2006)}]{sun:2006:ccm}
\bibinfo{author}{\bibfnamefont{C.-Y.} \bibnamefont{Sun}} \bibnamefont{and}
  \bibinfo{author}{\bibfnamefont{D.-H.} \bibnamefont{Zhang}}
  (\bibinfo{year}{2006}), \eprint{hep-th/0611101}.

\bibitem[{\citenamefont{Wesley et~al.}(2005)\citenamefont{Wesley, Steinhardt,
  and Turok}}]{wesley:2005:cct}
\bibinfo{author}{\bibfnamefont{D.~H.} \bibnamefont{Wesley}},
  \bibinfo{author}{\bibfnamefont{P.~J.} \bibnamefont{Steinhardt}},
  \bibnamefont{and} \bibinfo{author}{\bibfnamefont{N.}~\bibnamefont{Turok}},
  \bibinfo{journal}{Phys. Rev.} \textbf{\bibinfo{volume}{D72}},
  \bibinfo{pages}{063513} (\bibinfo{year}{2005}), \eprint{hep-th/0502108}.

\bibitem[{\citenamefont{Brandenberger et~al.}(2002)\citenamefont{Brandenberger,
  Easson, and Kimberly}}]{Brandenberger:2001:lpi}
\bibinfo{author}{\bibfnamefont{R.}~\bibnamefont{Brandenberger}},
  \bibinfo{author}{\bibfnamefont{D.~A.} \bibnamefont{Easson}},
  \bibnamefont{and} \bibinfo{author}{\bibfnamefont{D.}~\bibnamefont{Kimberly}},
  \bibinfo{journal}{Nucl. Phys.} \textbf{\bibinfo{volume}{B623}},
  \bibinfo{pages}{421} (\bibinfo{year}{2002}), \eprint{hep-th/0109165}.

\bibitem[{\citenamefont{Battefeld and Watson}(2006)}]{battefeld:2006:sgc}
\bibinfo{author}{\bibfnamefont{T.}~\bibnamefont{Battefeld}} \bibnamefont{and}
  \bibinfo{author}{\bibfnamefont{S.}~\bibnamefont{Watson}},
  \bibinfo{journal}{Rev. Mod. Phys.} \textbf{\bibinfo{volume}{78}},
  \bibinfo{pages}{435} (\bibinfo{year}{2006}), \eprint{hep-th/0510022}.

\bibitem[{\citenamefont{Brandenberger et~al.}(2006)\citenamefont{Brandenberger,
  Nayeri, Patil, and Vafa}}]{brandenberger:2006:sgc}
\bibinfo{author}{\bibfnamefont{R.~H.} \bibnamefont{Brandenberger}},
  \bibinfo{author}{\bibfnamefont{A.}~\bibnamefont{Nayeri}},
  \bibinfo{author}{\bibfnamefont{S.~P.} \bibnamefont{Patil}}, \bibnamefont{and}
  \bibinfo{author}{\bibfnamefont{C.}~\bibnamefont{Vafa}}
  (\bibinfo{year}{2006}), \eprint{hep-th/0608121}.

\bibitem[{\citenamefont{Brandenberger}(2007)}]{brandenberger:2007:sgc}
\bibinfo{author}{\bibfnamefont{R.}~\bibnamefont{Brandenberger}}
  (\bibinfo{year}{2007}), \eprint{hep-th/0702001}.

\bibitem[{\citenamefont{Brax et~al.}(2004)\citenamefont{Brax, van~de Bruck, and
  Davis}}]{brax:2004:bwc}
\bibinfo{author}{\bibfnamefont{P.}~\bibnamefont{Brax}},
  \bibinfo{author}{\bibfnamefont{C.}~\bibnamefont{van~de Bruck}},
  \bibnamefont{and} \bibinfo{author}{\bibfnamefont{A.-C.} \bibnamefont{Davis}},
  \bibinfo{journal}{Rept. Prog. Phys.} \textbf{\bibinfo{volume}{67}},
  \bibinfo{pages}{2183} (\bibinfo{year}{2004}), \eprint{hep-th/0404011}.

\bibitem[{\citenamefont{Campanelli et~al.}(2006)\citenamefont{Campanelli,
  Lousto, Marronetti, and Zlochower}}]{Campanelli:2005dd}
\bibinfo{author}{\bibfnamefont{M.}~\bibnamefont{Campanelli}},
  \bibinfo{author}{\bibfnamefont{C.~O.} \bibnamefont{Lousto}},
  \bibinfo{author}{\bibfnamefont{P.}~\bibnamefont{Marronetti}},
  \bibnamefont{and}
  \bibinfo{author}{\bibfnamefont{Y.}~\bibnamefont{Zlochower}},
  \bibinfo{journal}{Phys. Rev. Lett.} \textbf{\bibinfo{volume}{96}},
  \bibinfo{pages}{111101} (\bibinfo{year}{2006}), \eprint{gr-qc/0511048}.

\bibitem[{\citenamefont{Grumiller}()}]{danielpriv}
\bibinfo{author}{\bibfnamefont{D.}~\bibnamefont{Grumiller}},
  \emph{\bibinfo{title}{private communication}}.

\bibitem[{\citenamefont{Garcia et~al.}(2004)\citenamefont{Garcia, Hehl,
  Heinicke, and Macias}}]{Garcia:2003bw}
\bibinfo{author}{\bibfnamefont{A.}~\bibnamefont{Garcia}},
  \bibinfo{author}{\bibfnamefont{F.~W.} \bibnamefont{Hehl}},
  \bibinfo{author}{\bibfnamefont{C.}~\bibnamefont{Heinicke}}, \bibnamefont{and}
  \bibinfo{author}{\bibfnamefont{A.}~\bibnamefont{Macias}},
  \bibinfo{journal}{Class. Quant. Grav.} \textbf{\bibinfo{volume}{21}},
  \bibinfo{pages}{1099} (\bibinfo{year}{2004}), \eprint{gr-qc/0309008}.

\end{thebibliography}
\end{document}